# Appraising scattering theories for polycrystals of any symmetry using finite elements

Ming Huang[1], Stanislav I. Rokhlin[2] and Michael J. S. Lowe[1,*]

*[1]Department of Mechanical Engineering, Imperial College London, Exhibition Road, London SW7 2AZ, United Kingdom*

*[2]Department of Materials Science and Engineering, Edison Joining Technology Center, The Ohio State University, 1248 Arthur E. Adams Drive, Columbus, OH 43221, United States*

# Summary

This paper uses 3D grain-scale finite element (FE) simulations to appraise the classical scattering theory of plane longitudinal wave propagation in untextured polycrystals with statistically equiaxed grains belonging to the seven crystal symmetries. As revealed from the results of 10,390 materials, the classical theory has a linear relationship with the elastic scattering factor at the quasi-static velocity limit, whereas the reference FE and self-consistent (SC) results generally exhibit a quadratic relationship. As supported by the results of 90 materials, such order difference also extends to the attenuation and phase velocity, leading to larger differences between the classical theory and the FE results for more strongly scattering materials. Alternatively, two approximate models are proposed to achieve more accurate calculations by including an additional quadratic term. One model uses quadratic coefficients from quasi-static SC velocity fits and is thus symmetry-specific, while the other uses theoretically-determined coefficients and is valid for any individual material. These simple models generally deliver more accurate attenuation and phase velocity (particularly the second model) than the classical theory, especially for strongly scattering materials. However, the models are invalid for the attenuation of materials with negative quadratic coefficients.

# 1. Introduction

The scattering of elastic waves by the anisotropic grains of polycrystals is a classical problem of great practical importance in various fields, such as seismology [1,2] and non-destructive evaluation [3,4]. Studying this problem enables us to inform the grain structure of a polycrystal from scattering behaviours [5–8] and better discern the features of interest (e.g. fractures and voids) in a polycrystal by eliminating the influence of grain scattering [9,10]. A particular interest of this problem is in the scattering-induced attenuation and phase velocity variation of the propagating wave. Early studies on this topic focused on the individual Rayleigh [11], stochastic [12] and geometric [13] scattering regimes, as reviewed by Refs. [14,15]. Later on, Stanke and Kino [15] developed a unified theory valid for all scattering regimes by extending the Keller approximation [16,17] to the elastic wave problem. More recently, Weaver [18] formulated an equivalent theory by invoking the

* Author for correspondence (m.lowe@imperial.ac.uk).

Bourret approximation [19–21] (also known as the first-order smoothing approximation [18,19]) in the Dyson equation. These two equivalent treatments established the classical theory of grain scattering and spurred extensive later developments for polycrystals with differing grain structures and crystal symmetries [22–25].

Quantitative evaluation of the classical theory has been made possible by recent advancements in finite element (FE) simulations [26–34]. These studies used a grain-scale representation of polycrystals and performed FE simulations in a well-controlled condition, thus enabling an accurate description of elastic wave propagation and scattering within polycrystals [35]. They demonstrated a generally good agreement between the classical theory and simulation results for various cubic polycrystals with equiaxed [26–32] and elongated [33,34] grains.

The authors of this paper carried out the evaluation work to a deeper extent to see how the classical theory performs for strongly scattering cubic polycrystals [36,37] (equally important topic as for other inhomogeneous media [38,39]). In the low-frequency Rayleigh regime, we observed that the classical theory has a growing difference from 3D FE results as the elastic scattering factor (equivalent to the degree of inhomogeneity $\xi^2$ [15,40]) increases; their agreement remains good in the Rayleigh-stochastic transition region and the stochastic regime. As our analysis revealed, such difference arises because the classical theory can only predict attenuation and phase velocity variation to the linear order of the elastic scattering factor (due to its truncation of the solution to this order [15]), whereas the accurate FE results indicated that the true attenuation and velocity variation are of the quadratic order. Based on this finding, we suggested [36,37] an approximate model to account for the additional quadratic contribution, with the quadratic term coefficient obtained by fitting the model to the FE results. Although we determined the quadratic coefficient from cubic polycrystals with a specific microstructure, the resulting model exhibited excellent applicability to cubic polycrystals with various microstructures of different grain size distributions [36] and grain shapes [37], thus substantiating the generality of the approximate model.

The purpose of this paper is to conduct the FE evaluation work to the broadest ever extent to appraise the approximations of the classical theory for materials covering all seven crystal symmetries and to develop general approximate models to deliver more accurate attenuation and phase velocity calculations. First, we briefly summarise the classical theory and the 3D FE method in Sections 2 and 3, focusing on plane longitudinal waves in untextured polycrystals with statistically equiaxed grains. Then, we present the quasi-static velocity limit (essentially an elastostatic problem) results in Section 4, which extensively covers the classical self-consistent (SC) theory. We base our discussion on a vast set of materials (10,390 in total) belonging to the seven crystal symmetries, thus revealing a prominent general finding for the quasi-static velocity variation that is critical for the discussion in the next step. Next, we present the attenuation and phase velocity results in Section 5. We focus our discussion on developing two approximate models that include a quadratic term to achieve more accurate attenuation and phase velocity calculations. The first model is symmetry-specific, whereas the second is general and applicable to any polycrystal of arbitrary crystal symmetry. Finally, we conclude the paper with Section 6.

# 2. Classical theory

The classical theory considered here is the unified theory valid for all frequencies. We can formulate this theory in two different ways that lead to equivalent results, one by the perturbative Keller approximation [16] as used by Stanke and Kino [15], and another by the diagrammatic Bourret approximation [19–21] (also known as the first-order smoothing approximation [18,19]) of the Dyson equation as utilised by Weaver [18]. We use the latter approach and begin with the elastic wave equation [18]

$$\left\{\frac{\partial}{\partial x_i} c_{ijkl}(\mathbf{x}) \frac{\partial}{\partial x_l} + \rho\omega^2\delta_{jk}\right\} G_{k\alpha}(\mathbf{x},\mathbf{x}';\omega) = \delta_{j\alpha}\delta^3(\mathbf{x}-\mathbf{x}'), \tag{1}$$

where the Green's function $G_{k\alpha}(\mathbf{x},\mathbf{x}';\omega)$ describes the $k$-direction displacement response at $\mathbf{x}$ when the polycrystal is subject to an $\alpha$-direction unit impulse at $\mathbf{x}'$. $\omega = 2\pi f$ ($f$ is the frequency) is the angular frequency. $\delta_{jk}$ denotes the Kronecker delta and $\delta^3(\mathbf{x}-\mathbf{x}')$ the Dirac delta function. The mass density $\rho$ is constant throughout the single-phase polycrystal considered here. The elastic tensor $c_{ijkl}(\mathbf{x})$ is spatially varied and can be expressed as an average tensor $c_{ijkl}^0 = \langle c_{ijkl}(\mathbf{x})\rangle$ plus a fluctuation $\delta c_{ijkl}(\mathbf{x})$ about this average, namely $c_{ijkl}(\mathbf{x}) = c_{ijkl}^0 + \delta c_{ijkl}(\mathbf{x})$. The angle brackets denote the Voigt average [29], and the resulting tensor $c_{ijkl}^0$ describes the homogeneous, isotropic reference medium representing the polycrystal on the macro scale.

The solution to Equation (1) is intractable even for the scalar case, but the equation can be reduced to the Dyson integral equation when formulated for the mean Green's function [18]

$$\langle G_{k\alpha}(\mathbf{x},\mathbf{x}';\omega)\rangle = G_{k\alpha}^0(\mathbf{x},\mathbf{x}';\omega) + \iint G_{k\beta}^0(\mathbf{x},\mathbf{y};\omega) M_{\beta j}(\mathbf{y},\mathbf{z};\omega)\langle G_{j\alpha}(\mathbf{z},\mathbf{x}';\omega)\rangle d^3y d^3z, \tag{2}$$

where $G_{k\alpha}^0(\mathbf{x},\mathbf{x}';\omega)$ is the Green's function of the homogeneous reference medium. $M_{\beta j}(\mathbf{y},\mathbf{z};\omega)$ is the mass operator that accounts for the scattering events in the polycrystal. In the wave number and frequency domain, the mean Green's function can be given by the sum of three orthogonal modes [18]

$$\langle \mathbf{G}(\mathbf{k};\omega)\rangle = \sum_{M=1}^{3} \frac{\mathbf{u}_M \mathbf{u}_M}{\omega^2 - k^2 V_{0M}^2 - m_M(\mathbf{k};\omega)}, \tag{3}$$

with $\mathbf{k} = k\mathbf{p}$ ($k$ is the wave number and $\mathbf{p}$ the unit wave vector) and $\mathbf{u}_M$ representing the wave and polarisation vectors of the $M$ mode, respectively. $V_{0M}$ denotes the phase velocity in the reference medium. The zero of the denominator of the mean Green's function gives the dispersion equation for the effective wave number $k$ of the $M$ mode [33,34,41]

$$\omega^2 - k^2 V_{0M}^2 - m_M(\mathbf{k};\omega) = 0. \tag{4}$$

Since its exact solution remains intractable, the mass operator $m_M$ (now in the wave number domain) needs to be approximate by invoking the Bourret approximation [18]. The resulting expression for $m_M$ can be found in Refs. [32,36] for polycrystals with statistically equiaxed grains and in Refs. [33,34] for those with statistically elongated grains.

A vital element of the theory is the incorporation of the covariance of the elastic tensor in the mass operator. This covariance is given in the spatial domain by $\langle \delta c_{ijkl}(\mathbf{x}) \delta c_{\alpha\beta\gamma\delta}(\mathbf{x}') \rangle$, describing the statistical two-point correlation (TPC) between $\mathbf{x}$ and $\mathbf{x}'$ [32]. For a statistically homogeneous polycrystal, the covariance can be factored into elastic and spatial parts by $\langle \delta c_{ijkl}(\mathbf{x}) \delta c_{\alpha\beta\gamma\delta}(\mathbf{x}') \rangle = \langle \delta c_{ijkl} \delta c_{\alpha\beta\gamma\delta} \rangle w(\mathbf{x}-\mathbf{x}')$ [15,18]. The elastic part is solely determined by the elastic constants of the polycrystal. It is related to the second-order degree of inhomogeneity $\xi^2$ [15,40], which is alternatively represented by $Q_{M\to N}$ factors in this and our prior studies [32–34,36,41]. The spatial part $w(\mathbf{x}-\mathbf{x}')$ is the well-known spatial TPC function describing the probability of two points being in the same grain. The spatial TPC function is related to the size and shape of the grains, and it is scalar for statistically equiaxed grains, namely $w(\mathbf{x}-\mathbf{x}') = w(r)$ with $r = |\mathbf{x}-\mathbf{x}'|$, and direction-dependent for statistically elongated grains [34]. For comparison purposes, we will determine the spatial TPC function directly from 3D FE samples.

Upon supplying the above two covariance factors, we can solve Equation (4) using a variant of Newton's method. The solution for the wave number $k$ carries information about the attenuation and phase velocity of the propagating wave in its imaginary and real parts, i.e., $\alpha_M = \mathrm{Im}k$ and $V_M = \omega / \mathrm{Re}k$. Example attenuation and phase velocity curves are provided in Figure 1(b) for plane longitudinal waves ($M = \mathrm{L}$) in Inconel with statistically equiaxed grains. The curves cover the entire frequency range spanning from the low-frequency Rayleigh regime through the middle-frequency stochastic regime to the high-frequency geometric regime; the asymptotes of these three regimes are provided in the figure. We note that the curves will be nearly unchanged if the far-field Green's function is used in the mass operator (leading to the far-field approximation), but they will significantly deviate from the current ones at high frequencies if the Born approximation is invoked in the theory [32,36,41].

Due to their importance to this work, the Rayleigh attenuation and phase velocity asymptotes, at the limit of the wavelength being much larger than the average grain size, are provided here [32,36]

$$\alpha_M^{\mathrm{R}} = \frac{1}{2\pi} k_{0M}^4 V_{\mathrm{eff}}^{\mathrm{g}} \left( Q_{MM}^* + \frac{V_{0M}^3}{V_{0N}^3} Q_{M\to N} \right), \quad V_M^{\mathrm{R}} = \frac{V_{0M}}{1 + 2Q_{MM}^* + 2Q_{M\to N}}, \tag{5}$$

where $N \neq M$. $k_{0M}$ denotes the wave number of the wave $M$ in the reference medium. $V_{\mathrm{eff}}^{\mathrm{g}}$ is the effective grain volume defined by the volumetric integral of the TPC function [15,18,32]. $Q_{M\to N}$ is the elastic scattering factor describing the degree of inhomogeneity for the scattering from mode $M$ to $N$, while $Q_{MM}^*$ is an elastic factor introduced for simplifying the equation. We note that $Q_{M\to N}$ is the dominant elastic factor in both asymptotes for longitudinal waves ( $M = \mathrm{L}$ ) considered in this work, so we use it hereafter to describe the total degree of inhomogeneity.

The classical theory is necessarily approximate, as implied in the above steps. Among the various approximations, the most important ones are:

(1) The theory invokes the Bourret approximation. From the perturbative Keller approximation [16] point of view, this approximation truncates the solution to the second-order term in $\xi$ [15]. This implies that the theory is accurate only when $\xi^2$ (and thus the elastic anisotropy of the polycrystal) is small, and for this reason, we also call the theory the second-order approximation (SOA) [15]; while from the equivalent diagrammatic method point of view, the approximation limits the theory to account for only a subset of the scattering diagrams in the solution of the exact Dyson equation [19,20]. The neglected scattering events may be negligible for weakly scattering materials but become increasingly important as the elastic anisotropy of the polycrystal becomes stronger.

(2) The theory involves a major approximation by replacing a discrete polycrystal with a continuous random medium with fluctuating elastic tensor and statistically representing the polycrystal by the TPC function (covariance) [15,18–20]. This replacement is intuitively applicable to materials of weak elastic anisotropy but may introduce non-negligible errors for strongly scattering materials.

(3) The theory assumes the validity of factorising the above TPC function (covariance) into the elastic and spatial parts [15,18]. However, numerical studies supported this validity for polycrystals of macroscopic isotropy [42], so the factorisation approximation is negligible for the cases considered in this work.

(4) The theory considers the $n$ -point correlation function only to $n=2$ (the TPC function/covariance). The significance of the additional statistics on scattering remains unaddressed so far. Still, the fact of high order scattering diagrams depending on multipoint correlation functions [19,20] means that approximation exists due to the sole consideration of the $n=2$ statistic.

It is not yet known how these approximations affect the obtained solution, even for the scalar case, due to the lack of exact solutions. Therefore, numerical simulations are the only alternative at this point to evaluate the quality of the obtained solutions.

# 3. Finite element method

The 3D finite element (FE) method has been established recently to be powerful and accurate for the simulation of wave propagation in polycrystals [27,28,34,35]. Here we summarise our use of the method for this study for plane longitudinal waves in polycrystals with statistically equiaxed grains.

We begin with creating three numerical geometric samples with the Neper program [43] using Voronoi tessellation, as exemplified in Figure 1(a), with parameters in Table 1 of Ref. [36]. The samples are slab-shaped of fully bonded grains randomly packed in the sample domain. The average edge size of the grains is 0.5 mm by simply treating the grains as cubes (equivalent to an average radius of 0.31 mm by treating the grains as spheres) with a normal size distribution. As a more rigorous means, here we use the mean line intercept, which is related to the slope at the origin of the spatial TPC function [15,32,44], to characterise the average grain size, and the result is $a = 0.35$ mm. The thicknesses of the samples in the wave propagating z-direction are chosen to be around 10-wavelength (and ≥10 grains) long, and thus they range from 100 through 10 to 5 mm for the three samples that are for low-, middle- and high-frequency simulations, respectively. The widths of the samples in the transverse x- and y-directions are 12, 12 and 20 mm, so the samples contain 115200, 11520 and 16000 grains, far exceeding the statistical requirements set out by Refs. [45,46] for static homogenisation problems. The samples are discretised using structured 8-node linear elements with edge sizes of 0.05, 0.025 and 0.02 mm, respectively. This ensures that the grains are well represented after discretisation and the need for ~20 elements per wavelength is met to achieve a numerical error of ~0.1% [35]. We use these three geometric samples to simulate all materials in Table 1 that belong to the seven crystal symmetries.

To perform a simulation for a given material, the grains within the chosen sample are assigned with the same mass density and elastic constants of the material, but their crystallographic axes are uniformly randomly oriented, making the model macroscopically isotropic (untextured). Then, symmetry boundary conditions (SBCs) are defined for the four lateral surfaces as illustrated in Figure 1(a) by constraining the nodal displacements in the surface normal direction [28,35,36]. Next, a uniform force in the form of a three-cycle Hann-windowed toneburst is applied in the surface normal direction to all nodes on the $z = 0$ surface. The model is then solved in the time domain with the Pogo program [47] using a time-stepping scheme, with a time step of $\Delta t = 0.8h / V_{0\mathrm{L}}$ satisfying the Courant–Friedrichs–Lewy condition. Finally, we monitor the z-direction displacements during the time-stepping solution. An example result is provided in Figure 1(b) for Inconel simulated at a centre frequency of 1 MHz ($2k_{0\mathrm{L}}a \approx 1$) using the sample of 115200 grains (denoted N115200). As shown by the wavefield in panel A of the figure, the main wave pulse is partially scattered by the grains, leading to incoherent scattered waves in the space behind the main pulse. The incoherent scattered waves can be clearly observed in panel B in the signals monitored at individual nodes on the transmitting $z = 0$ and receiving $z = d_z$ surfaces; however, they are mostly cancelled out after averaging over all nodes (~60,000) on each surface, leaving only the unscattered coherent waves. Taking the main pulses (before the vertical marks in panel B) from the two coherent signals and transforming them into the frequency domain, we obtain the amplitude and phase spectra in panel C. Although these spectra cover a relatively broad frequency range, we only use the highlighted range for calculating the attenuation and phase velocity results shown in panel D. This is to achieve a high degree of numerical accuracy (error ~0.1%) for the results in the chosen frequency range; see our prior work [35] for the approach of selecting the appropriate frequency range.

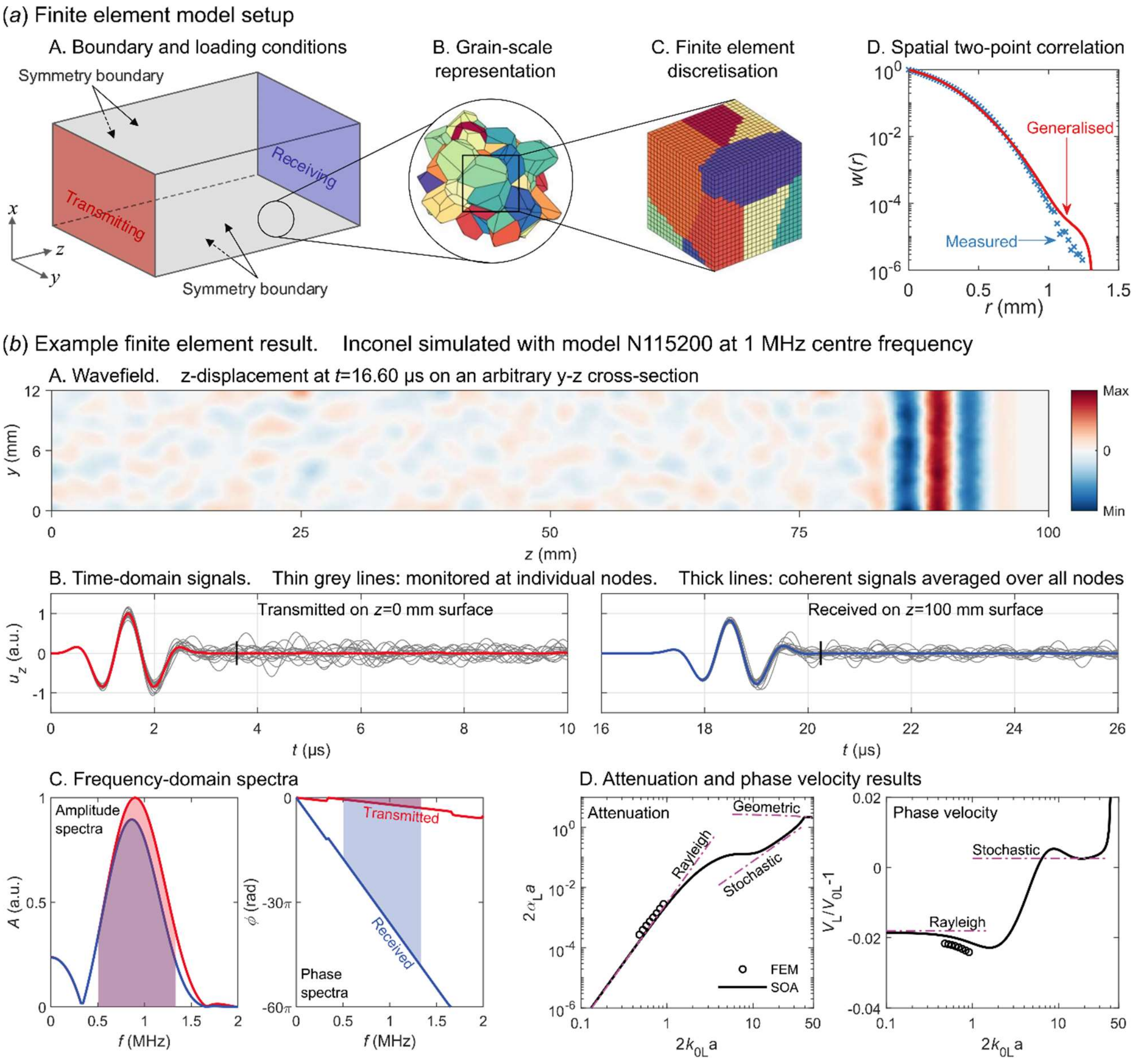

*Figure 1. FE model setup and example result. (a) FE model setup for simulating plane longitudinal waves in polycrystals in 3D. (b) Example result for plane longitudinal wave propagation in a single realisation of Inconel with statistically equiaxed grains, simulated with model N115200 using a centre frequency of 1 MHz for the transmitted toneburst. Panel A shows the wavefield at a given time on an arbitrary cross-section. Panel B shows the time-domain signals, highlighting the signals acquired at individual nodes and the coherent signals averaged over all nodes on the respective surfaces. Taking the main pulses from the coherent signals before the vertical marks in panel B and transforming them into the frequency domain leads to the amplitude and phase spectra in panel C. Panel D presents the normalised attenuation and phase velocity results calculated using the spectra in the highlighted frequency range in panel C. The normalisation factor $a$ is the mean line intercept representing the average grain size. The results are compared with the curves predicted by the classical SOA theory, with its Rayleigh, stochastic and geometric asymptotes given.*

We have now completed a single realisation of the desired simulation. To improve the statistical significance of the calculated attenuation and velocity results, we further perform multiple (15 in this work) realisations of the simulation and then take their average as the final result. The multiple realisations use the same spatial grain structures of the chosen sample but reshuffle the crystallographic orientations of the grains. Such reshuffling essentially creates different ensembles of grains, equivalent to the approach by generating different spatial grain structures for each realisation [28]. To obtain results for a broad frequency range, the centre frequency of the transmitted toneburst varies between 0.5 and 20 MHz. For each centre frequency, an optimal geometric sample that maximises the accuracy of the simulation is utilised [35], and multiple realisations of the simulation are conducted using the chosen model. Altogether, attenuation and velocity results of similar numerical accuracy and statistical significance are obtained for the simulated material over a broad frequency, ranging from a normalised frequency $2k_{0\mathrm{L}}a$ of around 0.5 to 15.

*Table 1. Polycrystalline materials for FE simulation. Equivalent anisotropy index* $A^{\mathrm{eq}}$ [48] ( $A^{\mathrm{eq}} = A$ *for cubic materials), universal anisotropy index* $A^{\mathrm{U}}$ [49], *log-Euclidean anisotropy index* $A^{\mathrm{L}}$ [50], *elastic scattering factor* $Q_{\mathrm{L}\to\mathrm{T}}$. *Elastic constants and densities can be found in the cited references. The materials highlighted in bold are simulated in a broad frequency range, while the rest are simulated only at low frequencies at* $2k_{0\mathrm{L}}a \approx 1$. *The Orthorhombic* row gives two extreme materials.*

| | $A^{\mathrm{eq}} < 1$ | | | | | $A^{\mathrm{eq}} > 1$ | | | | |
|---|---|---|---|---|---|---|---|---|---|---|
| | Materials | $A^{\mathrm{eq}}$ | $A^{\mathrm{U}}$ | $A^{\mathrm{L}}$ | $Q_{\mathrm{L}\to\mathrm{T}}$ | Materials | $A^{\mathrm{eq}}$ | $A^{\mathrm{U}}$ | $A^{\mathrm{L}}$ | $Q_{\mathrm{L}\to\mathrm{T}}$ |
| Cubic | **Cr** [51] | 0.71 | 0.14 | 0.06 | 1.40E-03 | **Al** [36] | 1.24 | 0.05 | 0.02 | 3.34E-04 |
| | Nb [51] | 0.49 | 0.63 | 0.26 | 3.13E-03 | A=1.5 [36] | 1.52 | 0.21 | 0.09 | 1.43E-03 |
| | **RbF** [52] | 0.45 | 0.80 | 0.33 | 6.44E-03 | A=1.8 [36] | 1.83 | 0.45 | 0.19 | 2.79E-03 |
| | $As_1$ [53] | 0.39 | 1.17 | 0.47 | 9.98E-03 | A=2.4 [36] | 2.45 | 1.03 | 0.42 | 5.48E-03 |
| | RbCl [52] | 0.31 | 1.82 | 0.69 | 1.38E-02 | Cu [36] | 3.14 | 1.75 | 0.67 | 7.19E-03 |
| | RbBr [52] | 0.28 | 2.19 | 0.81 | 1.67E-02 | **Inconel** [36] | 2.83 | 1.42 | 0.56 | 7.59E-03 |
| | **RbI** [52] | 0.25 | 2.64 | 0.95 | 1.86E-02 | A=5.0 [36] | 5.00 | 3.84 | 1.27 | 1.27E-02 |
| | $Dy_1S_1$ [53] | 0.18 | 4.35 | 1.40 | 2.50E-02 | **Li** [36] | 9.14 | 8.70 | 2.25 | 1.87E-02 |
| Hexagonal | $\alpha$-Ti [54] | 0.68 | 0.18 | 0.08 | 1.22E-03 | **$\alpha$-Be** [51] | 1.24 | 0.05 | 0.02 | 7.10E-04 |
| | $\alpha$-Co [51] | 0.67 | 0.20 | 0.09 | 1.82E-03 | $Nb_4N_4$ [53] | 1.63 | 0.29 | 0.12 | 2.46E-03 |
| | $\alpha$-Tl [51] | 0.30 | 1.98 | 0.75 | 5.72E-03 | $Mn_2Sb_2$ [53] | 2.09 | 0.68 | 0.28 | 6.52E-03 |
| | $Mn_2Bi_2$ [53] | 0.63 | 0.27 | 0.14 | 9.85E-03 | **Zn** [54] | 3.06 | 1.67 | 0.60 | 9.97E-03 |
| | $Ba_2$ [53] | 0.28 | 2.29 | 0.84 | 1.76E-02 | **$Sr_4Si_2$** [53] | 6.15 | 5.17 | 1.58 | 1.83E-02 |
| | $Be_2Se_2$ [53] | 0.11 | 8.78 | 2.27 | 2.50E-02 | $Al_2Cu_2O_6$ [53] | 165.09 | 195.71 | 8.26 | 2.34E-02 |
| Tetragonal | $Mg_1Al_3$ [53] | 0.72 | 0.13 | 0.05 | 1.02E-03 | **$Sm_4O_2$** [53] | 1.23 | 0.05 | 0.02 | 9.96E-04 |
| | PDP [55] | 0.39 | 1.12 | 0.45 | 4.59E-03 | Sn [55] | 2.76 | 1.34 | 0.53 | 3.39E-03 |
| | $K_2N_6$ [53] | 0.50 | 0.60 | 0.24 | 6.30E-03 | **Rutile** [55] | 2.76 | 1.34 | 0.52 | 8.70E-03 |
| | $Zr_1H_2$ [53] | 0.29 | 2.09 | 0.77 | 9.92E-03 | NSH [55] | 2.96 | 1.55 | 0.58 | 1.27E-02 |
| | RDP [55] | 0.21 | 3.48 | 1.18 | 2.02E-02 | **TO** [55] | 11.34 | 11.32 | 2.64 | 1.83E-02 |
| | $Li_2C_1N_2$ [53] | 0.26 | 2.53 | 0.88 | 2.47E-02 | $Tl_2Cu_1F_4$ [53] | 6.15 | 5.17 | 1.56 | 2.67E-02 |
| Trigonal | DSB [55] | 0.78 | 0.08 | 0.03 | 1.27E-03 | **$Na_6P_2S_2O_6$** [53] | 1.34 | 0.11 | 0.05 | 9.83E-04 |
| | $In_4O_6$ [53] | 0.71 | 0.15 | 0.06 | 2.56E-03 | $LiTaO_3$ [55] | 1.60 | 0.27 | 0.11 | 3.30E-03 |
| | Tourmaline [55] | 0.53 | 0.51 | 0.20 | 6.69E-03 | **$SiO_2$** [55] | 2.26 | 0.84 | 0.34 | 9.58E-03 |
| | Bi [55] | 0.31 | 1.79 | 0.66 | 1.29E-02 | Calcite [55] | 3.31 | 1.94 | 0.71 | 1.46E-02 |
| | $Cr_2Ag_2O_4$ [53] | 0.16 | 5.15 | 1.57 | 1.97E-02 | **Sb** [55] | 4.06 | 2.77 | 0.95 | 2.29E-02 |
| | $Mg_1Al_2H_8$ [53] | 0.27 | 2.37 | 0.82 | 2.48E-02 | GASH [55] | 3.28 | 1.90 | 0.71 | 2.84E-02 |
| Orthorhombic | $Co_2Se_4$ [53] | 0.74 | 0.11 | 0.05 | 9.93E-04 | **Enstatite** [56] | 1.37 | 0.12 | 0.05 | 1.40E-03 |
| | $Ni_2SiO_4$ [56] | 0.71 | 0.15 | 0.06 | 2.31E-03 | $FeSiO_3$ [56] | 1.51 | 0.20 | 0.08 | 2.70E-03 |
| | $Fe_2SiO_4$ [56] | 0.53 | 0.51 | 0.21 | 5.37E-03 | **U** [57] | 1.87 | 0.49 | 0.20 | 6.03E-03 |
| | $Ca_2Ag_4$ [53] | 0.51 | 0.58 | 0.23 | 1.02E-02 | $Sn_4Pd_4$ [53] | 3.49 | 2.13 | 0.78 | 9.85E-03 |
| | $Mo_4O_{10}$ [53] | 0.37 | 1.28 | 0.47 | 1.76E-02 | **$Li_4Nb_4N_8$** [53] | 3.10 | 1.71 | 0.62 | 1.77E-02 |
| | $Na_2Cu_1O_2$ [53] | 0.19 | 4.10 | 1.33 | 2.55E-02 | $Te_8O_{16}$ [53] | 3.23 | 1.85 | 0.66 | 2.43E-02 |
| Orthorhombic* | $Na_2U_1O_4$ [53] | 0.26 | 2.50 | 0.86 | 4.00E-02 | $Sr_1Mg_6Ga_1$ [53] | 325.36 | 388.04 | 10.24 | 3.46E-02 |
| Monoclinic | $\alpha/\beta$ Ti [58] | 0.65 | 0.22 | 0.10 | 1.52E-03 | **$Ca_4V_4O_{12}$** [53] | 1.33 | 0.10 | 0.04 | 9.28E-04 |
| | $NaAlSi_2O_6$ [56] | 0.62 | 0.28 | 0.11 | 2.80E-03 | $Al_8Mo_3$ [53] | 1.60 | 0.27 | 0.12 | 2.54E-03 |
| | $LiAlSi_2O_6$ [56] | 0.53 | 0.51 | 0.20 | 7.40E-03 | **$CaMgSi_2O_6$** [56] | 1.74 | 0.38 | 0.15 | 6.63E-03 |
| | Coesite [56] | 0.41 | 1.01 | 0.39 | 1.08E-02 | $Ba_2C_4$ [53] | 3.71 | 2.38 | 0.86 | 1.09E-02 |
| | Gypsum [56] | 0.33 | 1.59 | 0.60 | 1.56E-02 | **$Li_2Mg_4$** [53] | 7.60 | 6.87 | 1.93 | 1.82E-02 |
| | Muscovite [56] | 0.23 | 3.20 | 1.05 | 2.90E-02 | $Li_2Cu_2F_8$ [53] | 3.39 | 2.02 | 0.71 | 2.60E-02 |
| Triclinic | $Ba_9Nb_{10}O_{30}$ [53] | 0.72 | 0.14 | 0.06 | 1.20E-03 | **$Ce_8Y_8O_{28}$** [53] | 1.47 | 0.18 | 0.08 | 1.42E-03 |
| | $Ca_8Zr_2N_8$ [53] | 0.49 | 0.64 | 0.27 | 3.96E-03 | $Sr_4Co_4O_{10}$ [53] | 1.84 | 0.46 | 0.18 | 3.93E-03 |
| | $Al_4C_3$ [53] | 0.38 | 1.19 | 0.46 | 1.09E-02 | **CSP** [32] | 2.37 | 0.95 | 0.37 | 7.19E-03 |
| | $Rb_4Nb_4O_{12}$ [53] | 0.35 | 1.43 | 0.52 | 1.78E-02 | $Ca_6Al_6N_{10}$ [53] | 2.78 | 1.36 | 0.53 | 9.78E-03 |
| | Albite [59] | 0.29 | 2.09 | 0.74 | 2.36E-02 | **$Mo_4O_{12}$** [53] | 3.96 | 2.65 | 0.93 | 1.74E-02 |
| | $Co_1H_2O_2$ [53] | 0.12 | 7.70 | 2.03 | 4.49E-02 | $Zr_1Cu_1F_6$ [53] | 4.70 | 3.50 | 1.13 | 3.08E-02 |

In addition to their high degree of numerical accuracy and statistical significance, we highlight that the simulation results account for all possible scattering events in the simulated polycrystal. Therefore, the simulation results are suited for evaluating the approximations of the classical theory. The evaluation relies on putting the exact statistics of the simulated polycrystal into the theory. As aforementioned, the elastic TPC function required by the theory is calculated from the elastic constants in a Voigt-average sense, which is the same setting for the simulations. The required spatial TPC function is directly measured from the numerical geometric samples, shown as the points in panel D of Figure 1(a); it is further represented by a generalised mathematical function for direct incorporation into the theory, shown as the curve in the same figure. The generalised TPC function can be found in Ref. [36].

By slightly changing its boundary and loading conditions, the above FE method is also used in this work to calculate the phase velocity of polycrystals at the quasi-static limit to a high degree of accuracy (error ~0.01%); see details in Refs. [29,32].

# 4. Quasi-static velocity limit

We begin our discussion with the quasi-static velocity limit. This limit is essentially a static problem, but we treat it as an elastodynamic one at the limit of $f \to 0$ (i.e., $\lambda \to \infty$) such that the concept of phase velocity still holds. In this sense, we can extend the insights gained here later to the elastodynamic problem. Since we address longitudinal waves in this work, the discussion here is for the longitudinal quasi-static velocity limit. This velocity limit, denoted $V_{\mathrm{L}}$, is related to the effective elastic constant $C_{11}$ by $V_{\mathrm{L}} = \sqrt{C_{11} / \rho}$.

The quasi-static velocity results are shown in Figure 2 for materials of different symmetries. (a) presents the normalised variation in quasi-static velocity from the Voigt velocity, $V_{\mathrm{L}} / V_{0\mathrm{L}} - 1$, calculated using the FE method, the Rayleigh asymptote of the classical SOA theory, Equation (5), and the self-consistent (SC) theory [60,61]. (b) displays the relative difference between these three sets of results. Notice that the Hashin-Shtrikman bounds [62,63] (or the even tighter Hill-Mandel bounds [64]) are not provided here, and readers are referred to our prior work [36] for those for cubic materials.

The results are given for two groups of materials. One group is the 90 materials listed in Table 1, and the other group involves 10300 materials obtained from various sources [51,53,55,56,65]. The number of materials belonging to each of the seven crystal symmetries is summarised in Table 2 (and Table 3). Most of the materials are the compounds obtained from the Materials Project [53,65], generated from first-principle calculations using the density-functional theory. We retrieved all compounds with available elastic tensors from the Materials Project and then omitted those that (1) do not satisfy the elastic stability conditions [66] or (2) have the elastic scattering factor $Q_{\mathrm{L}\to\mathrm{T}}$ greater than 0.05. The choice of materials with $Q_{\mathrm{L}\to\mathrm{T}} \leq 0.05$ is to maintain the quality of figure plots but without affecting the conclusion made from the plots. Both groups of materials are provided with SOA and SC results, but only the first group of materials is provided with FE modelling (FEM) results. Each FEM point in the figure is the average of 15 realisations of the respective material. We emphasise that a single realisation has already achieved a high degree of statistical significance by containing a large number of grains in the FE sample that far exceed the number suggested by Refs. [45,46]; yet, the statistical significance is further improved by using a combination of 15 realisations.

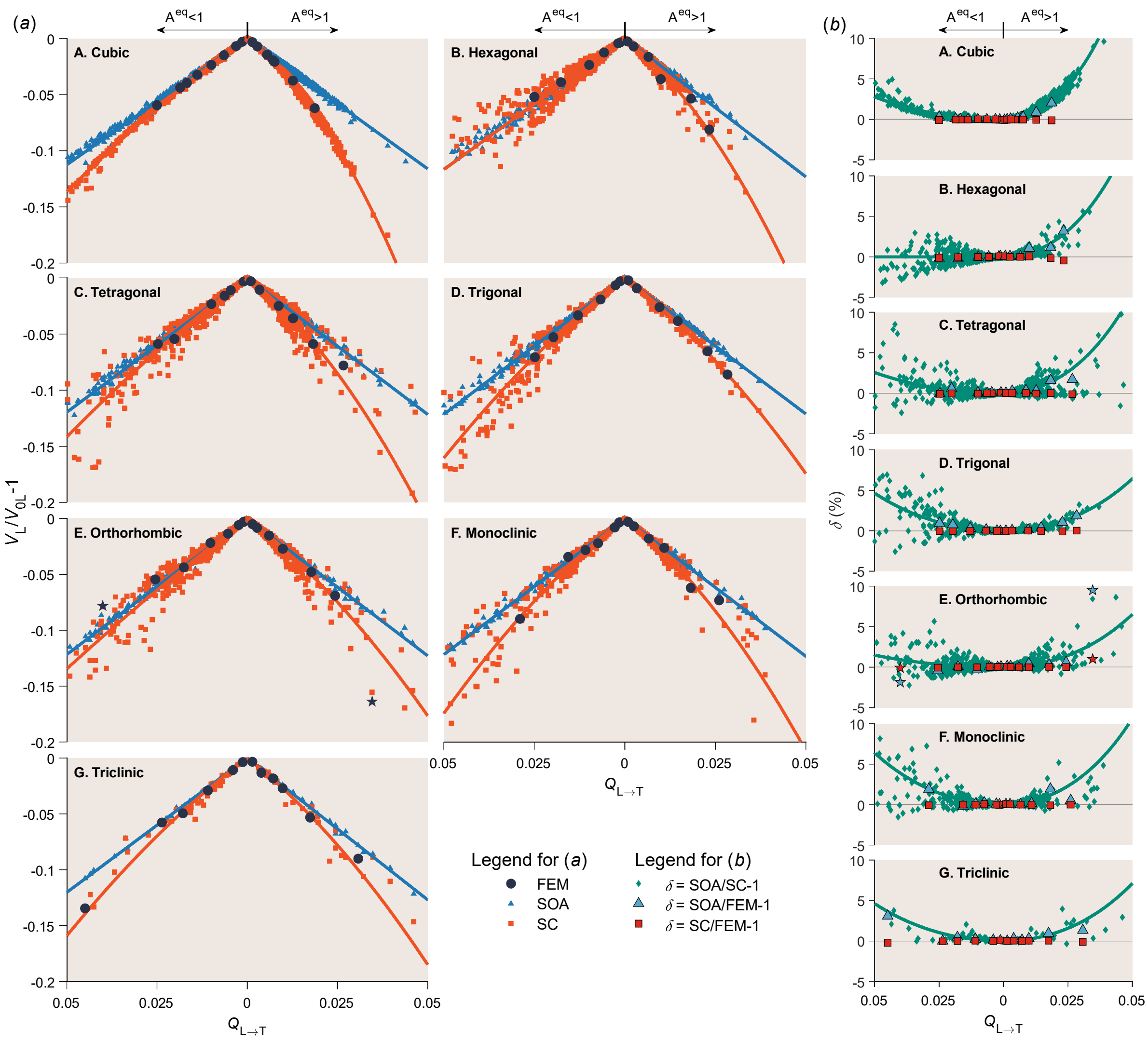

*Figure 2. Phase velocity at the quasi-static limit. (a) Quasi-static velocity variation against elastic scattering factor $Q_{\mathrm{L}\to\mathrm{T}}$ for materials of different symmetries with statistically equiaxed grains. The points are SOA, SC and FEM results for individual materials, while the lines are linear and quadratic fits to the SOA and SC points, respectively. (b) Relative difference in quasi-static velocity. The points are the relative difference between the points shown in (a), while each line is the relative difference between the respective SOA and SC fits in (a). Note that the star points in the E panels are for the two extreme materials in Table 1 that will be examined in Section 5.*

The results are plotted against the elastic scattering factor $Q_{\mathrm{L}\to\mathrm{T}}$, representing the elastic anisotropy (or the degree of inhomogeneity). Here $Q_{\mathrm{L}\to\mathrm{T}}$ is used for two reasons. First, $Q_{\mathrm{L}\to\mathrm{T}}$ appears explicitly in the Rayleigh asymptote of the SOA theory, Equation (5). Second, we identify that $Q_{\mathrm{L}\to\mathrm{T}}$ has a better monotonic correlation with the accurate SC results than all other anisotropy indices, including the equivalent $A^{\mathrm{eq}}$ [48], universal $A^{\mathrm{U}}$ [49] and log-Euclidean $A^{\mathrm{L}}$ [50] indices. The monotonic correlation of the SC points in Figure 2(a) with each of $A^{eq}$, $A^{U}$, $A^{L}$ and $Q_{L\to T}$ is quantified using Spearman's coefficient of rank correlation [67]. The correlation results are provided in Table 2 for each of the two branches of individual symmetries. The results show that $Q_{L\to T}$ has a near-perfect correlation with the SC points, with a coefficient of around 0.99-1 for all cases, whereas the three anisotropy indices have a less satisfactory correlation, with a coefficient of 0.90-0.98. For this reason, $Q_{L\to T}$ is used throughout this work to represent the elastic anisotropy of polycrystals.

*Table 2. Spearman's coefficient of rank correlation* [67]. *Each result quantifies the monotonic correlation between the self-consistent results* $V_{L}^{SC}/V_{0L}-1$ *of* $N$ *materials and each of the anisotropy indices* $A^{eq}$, $A^{U}$, $A^{L}$ *and the elastic scattering factor* $Q_{L\to T}$ *. Coefficients of 1 and −1 represent perfect positive and negative correlation; 0 means uncorrelated.*

| | $A^{eq}<1$ | | | | | $A^{eq}>1$ | | | | |
|---|---|---|---|---|---|---|---|---|---|---|
| | $N$ | $A^{eq}$ | $A^{U}$ | $A^{L}$ | $Q_{L\to T}$ | $N$ | $A^{eq}$ | $A^{U}$ | $A^{L}$ | $Q_{L\to T}$ |
| Cubic | 1652 | 0.9805 | -0.9805 | -0.9805 | -0.9997 | 2782 | -0.9709 | -0.9709 | -0.9709 | -0.9998 |
| Hexagonal | 867 | 0.9464 | -0.9464 | -0.9443 | -0.9955 | 597 | -0.9326 | -0.9326 | -0.9312 | -0.9947 |
| Tetragonal | 650 | 0.9181 | -0.9181 | -0.9111 | -0.9925 | 942 | -0.9313 | -0.9313 | -0.9243 | -0.9900 |
| Trigonal | 505 | 0.9578 | -0.9578 | -0.9525 | -0.9962 | 202 | -0.9696 | -0.9696 | -0.9667 | -0.9970 |
| Orthorhombic | 727 | 0.9206 | -0.9206 | -0.9146 | -0.9945 | 735 | -0.9392 | -0.9392 | -0.9322 | -0.9948 |
| Monoclinic | 373 | 0.9236 | -0.9236 | -0.9160 | -0.9936 | 268 | -0.9374 | -0.9374 | -0.9292 | -0.9943 |
| Triclinic | 51 | 0.9422 | -0.9422 | -0.9262 | -0.9893 | 39 | -0.9028 | -0.9028 | -0.8955 | -0.9864 |

The results are separated into $A^{eq}<1$ and $A^{eq}>1$ branches for each of the seven crystal symmetries. Such separation is based on our earlier observation for cubic materials that exhibit contrasting dependences on $Q_{L\to T}$ between $A<1$ and $A>1$ cases [36]; note that the Zener anisotropy index $A$ equals $A^{eq}$ for cubic materials. To evaluate the necessity of such separation, the statistical equality between the two branches of each symmetry is tested using Fisher's Z transformation [68]. The test leads to Z statistics of -22.83, 3.44, -2.89, 8.26, -0.09, -0.29 and -0.06 for the seven symmetries. The critical Z value for rejecting equality is 2.58 at a significance level of $\alpha=0.01$ [68]. Therefore, the two branches for the cubic, hexagonal, tetragonal and trigonal materials are statistically unequal and thus need to be separated. By contrast, those for the orthorhombic, monoclinic and triclinic materials are favoured for not being separated, though this may not be conclusive for triclinic materials considering their small sample sizes of <100. For consistency, we separate all symmetries into two branches in this work. We point out that sorting materials into two branches and then ordering materials by $Q_{L\to T}$ might be an interesting new way for systematically ordering materials of different properties, which is a topic that received extensive attention, see e.g. [69–71]. In particular, cubic materials are perfectly sorted into two branches, with each tightly following a monotonic curve of $Q_{L\to T}$; this is also true for the $A^{eq}>1$ branch of trigonal materials.

The SC results have an excellent agreement with the FEM results for all seven symmetries, showing a relative difference at the level of 0.1% (max 1%, for orthorhombic $Sr_1Mg_6Ga_1$). In addition to this evidence, the exceptionally high degree of accuracy of the SC theory was also extensively supported by prior FE results [32,36,72]; the underlying reason for this is that the SC theory satisfies the continuity of stress and strain throughout the polycrystal [60,61]. Therefore, in addition to FE calculations, we use the SC results in this paper as the reference to appraise the classical SOA theory at the quasi-static limit (this allows us to reduce the amount of very computationally-intensive FE calculations).

The SOA theory generally shows a growing deviation from the SC (and FEM) as $Q_{\mathrm{L}\to\mathrm{T}}$ increases. In-depth analysis reveals that the SOA quasi-static velocity only exhibits a linear relationship with $Q_{\mathrm{L}\to\mathrm{T}}$ while the accurate SC (and FEM) results mostly have a quadratic relationship. To illustrate this, we have generated linear and quadratic fits for the two cases, which are detailed in Table 3 and shown as lines in Figure 2(a). The fits are generated in a multiobjective sense by simultaneously fitting $V_{\mathrm{L}}/V_{0\mathrm{L}}-1=-2Q_{\mathrm{L}\to\mathrm{T}}(l+2qQ_{\mathrm{L}\to\mathrm{T}})$ to both the SOA and SC results, but with the constraint of $q=0$ for fitting to the SOA results. $l$ and $q$ are the linear and quadratic coefficients.

*Table 3. Fitting results. Each fitting is performed in a multiobjective sense by fitting $V_{\mathrm{L}}/V_{0\mathrm{L}}-1=-2Q_{\mathrm{L}\to\mathrm{T}}(l+2qQ_{\mathrm{L}\to\mathrm{T}})$ to both the SOA and SC results of $N$ materials, with $q=0$ for fitting to the SOA results. The linear $l$ and quadratic $q$ coefficients are given with standard deviations. The goodness of fit is quantified by $R^2$, with a value of 1 representing a perfect fit.*

| | $A^{\mathrm{eq}}<1$ | | | | $A^{\mathrm{eq}}>1$ | | | |
|---|---|---|---|---|---|---|---|---|
| | $N$ | $q$ | $l$ | $R^2$ | $N$ | $q$ | $l$ | $R^2$ |
| Cubic | 1652 | 2.40+0.017 | 1.12+0.001 | 0.998 | 2782 | 15.26+0.034 | 1.16+0.001 | 0.998 |
| Hexagonal | 867 | 0.00+0.137 | 1.17+0.003 | 0.929 | 597 | 11.80+0.280 | 1.23+0.003 | 0.967 |
| Tetragonal | 650 | 2.20+0.157 | 1.20+0.003 | 0.936 | 942 | 9.71+0.236 | 1.22+0.002 | 0.942 |
| Trigonal | 505 | 3.90+0.185 | 1.22+0.003 | 0.940 | 202 | 5.31+0.156 | 1.21+0.004 | 0.990 |
| Orthorhombic | 727 | 1.29+0.137 | 1.22+0.002 | 0.947 | 735 | 5.37+0.206 | 1.23+0.002 | 0.945 |
| Monoclinic | 373 | 5.28+0.210 | 1.22+0.003 | 0.953 | 268 | 8.45+0.366 | 1.23+0.004 | 0.948 |
| Triclinic | 51 | 3.89+0.361 | 1.20+0.007 | 0.975 | 39 | 5.81+0.639 | 1.27+0.008 | 0.964 |

The fits, which are based upon large sets of materials, match both the SOA and SC results well across all seven symmetries, with a goodness of fit of $R^2\geq 0.93$, thus corroborating their linear and quadratic relationships to $Q_{\mathrm{L}\to\mathrm{T}}$. The SC fits have small quadratic coefficients for the $A^{\mathrm{eq}}<1$ branches of hexagonal, tetragonal and orthorhombic materials. This means the related SC datasets behave predominantly linearly, reflecting a good agreement between the SOA and SC points for these cases. The linear coefficients of the SC fits remain nearly the same between the $A^{\mathrm{eq}}<1$ and $A^{\mathrm{eq}}>1$ branches of each symmetry. The quadratic coefficients only differ slightly between the two branches for the orthorhombic, monoclinic and triclinic symmetries, while the difference is more pronounced for the cubic, hexagonal, tetragonal and trigonal symmetries. These are consistent with the above equality test results.

Most importantly, the good agreement between the fits and the quasi-static results leads us to postulate that we can describe the quasi-static velocity variation well up to the second order of $Q_{\mathrm{L}\to\mathrm{T}}$, among which:

(1) The linear part is fully accounted for by the SOA theory, which is in line with its formalism by truncating the solution to the linear order of $Q_{\mathrm{L}\to\mathrm{T}}$ (equivalently, $\xi^2$ [15,40]).

(2) The quadratic part is solely attributed to the SC-SOA difference.

This is especially true for cubic materials and the $A^{\mathrm{eq}} > 1$ branch of trigonal materials, whose SOA and SC fits nearly perfectly match the respective points. For other cases, the SOA points align excellently with the fits, thus satisfying the above linear part inference; however, the SC points spread out substantially about the fits, especially at large $Q_{\mathrm{L}\to\mathrm{T}}$. Therefore, to account for such spreading, we propose that the above linear and quadratic inferences also hold for individual materials, meaning that we can partition the quasi-static velocity variation of any material into two parts, with the SOA theory describing the linear part and the SC-SOA difference representing the quadratic part. We will revisit this later in §0.

We point out that the quasi-static velocity is continuously connected between the two branches at $A^{\mathrm{eq}} = 1$ (isotropy), as shown in Figure 2(a) across all seven symmetries. This is especially clear from Figure 2(b) for the relative difference between the SOA and SC results that show a smooth transition. This is highlighted by the lines in Figure 2(b) that are the relative difference between the SOA and SC fits.

# 5. Attenuation and phase velocity

Now we investigate the elastodynamic problem, discussing the attenuation and phase velocity of plane longitudinal waves in polycrystals of different crystal symmetries.

Figure 3 shows the normalised attenuation and phase velocity variation versus the normalised frequency for the materials highlighted in Table 1. These include three cubic materials with $A^{\mathrm{eq}} < 1$, and three materials with $A^{\mathrm{eq}} > 1$ for each of the seven symmetries; the selected materials have comparable $Q_{\mathrm{L}\to\mathrm{T}}$ values from one case to another. The FEM points are obtained by taking the average of 15 realisations, and their small error bars are not shown. The SOA curves are calculated using Equation (4) by incorporating the TPC function of the FE samples.

The SOA curves have an excellent agreement with the FEM points for materials of small $Q_{\mathrm{L}\to\mathrm{T}}$ (with the lowest attenuation curves in each panel of Figure 3 and phase velocity variation curves at the top), thus cross-validating the correctness of both results. However, in most cases, the SOA curves start to deviate from the FEM points as $Q_{\mathrm{L}\to\mathrm{T}}$ increases. Notably, such deviation is pronounced in the low-frequency range, whereas it is small and barely shows $Q_{\mathrm{L}\to\mathrm{T}}$ dependence at high frequencies.

Therefore, our focus will be on the low-frequency range, attempting to introduce two approximate models with improved accuracy in this frequency range, especially for materials of large $Q_{\mathrm{L}\to\mathrm{T}}$. The first model is symmetry-specific, established for each branch of the seven crystal symmetries using the quasi-static SC fit discussed in the preceding section. By contrast, the second model is general and direct, initiated for individual materials using their respective quasi-static SOA and SC results. Hence, we name them as fitted (FAM) and direct (DAM) approximate models.

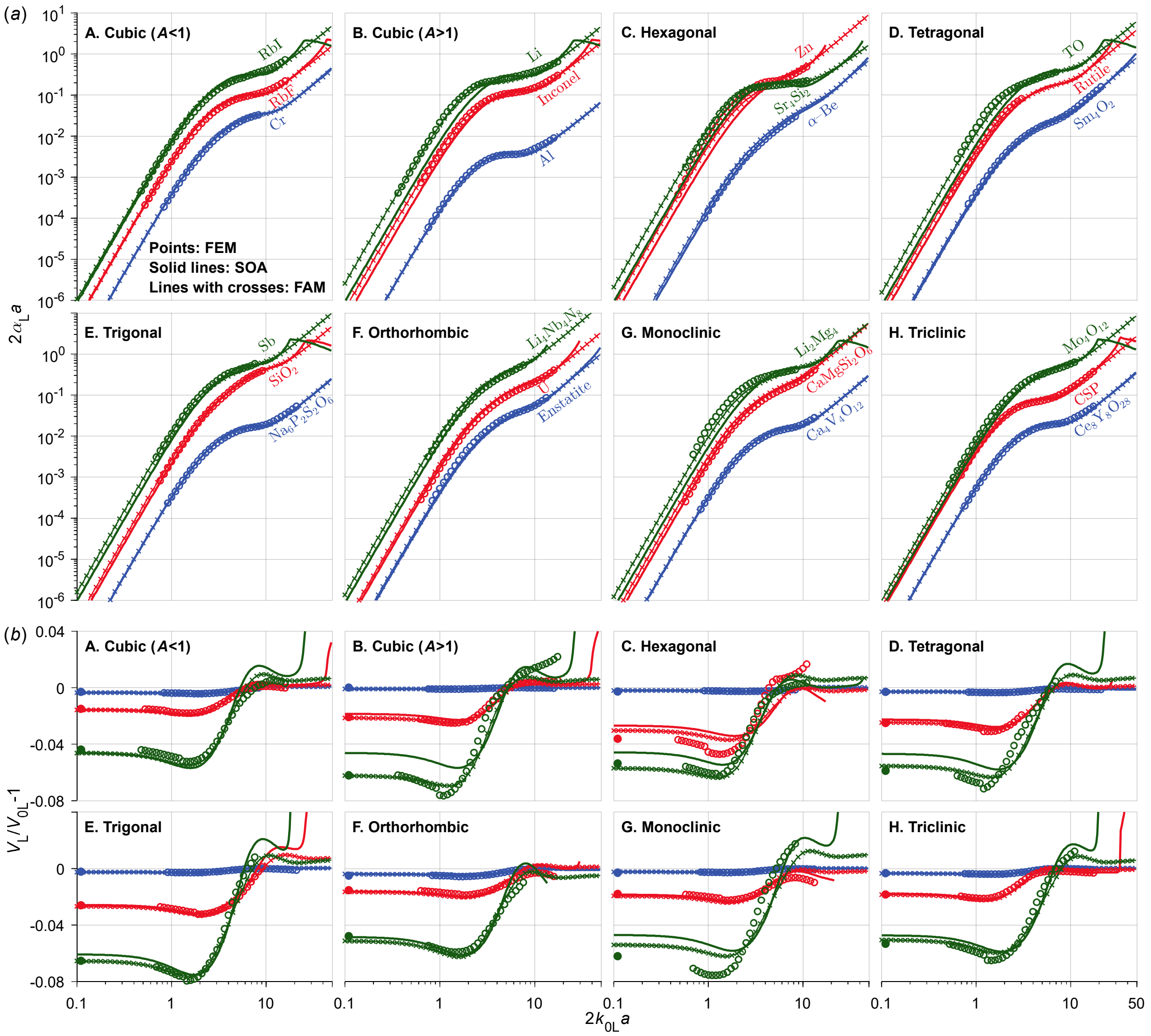

*Figure 3. Attenuation and phase velocity of plane longitudinal waves in materials of different crystal symmetries, comparing the FEM, SOA and FAM results. (a) shows the normalised attenuation and (b) the respective phase velocity variation against the normalised frequency* $2k_{0L}a$ *. All FEM points are obtained by taking the average of 15 realisations; the tiny error bars are not shown. The quasi-static FEM results (leftmost solid points in (b)) are taken from Figure 2. The SOA curves are predicted using Equation (4) by incorporating the TPC function of the FE samples. The FAM curves are obtained from Equations (6) and (7) using the fitted quadratic coefficients in Table 3.*

## (a) Fitted approximate model (FAM)

The idea of the FAM comes from our prior work [36]. For cubic materials, we demonstrated that the classical SOA theory only predicts attenuation and phase velocity variation to the first order of $Q_{L\to T}$, whereas the accurate FEM results are of the second order. Such order difference is consistent with the above quasi-static SOA and the quasi-static FEM and SC results. Based on this observation, as in our prior work [36], we include a quadratic term of $Q_{L\to T}$ in the FAM model; without this quadratic term, this model would degenerate to the SOA model (which invokes the Born approximation and uses the far-field Green's function in the mass operator). This model is given by

$$\alpha_{\mathrm{L}} = \sum_i A_i \frac{4Q_{\mathrm{L}\to\mathrm{L}} k_{0\mathrm{L}} (k_{0\mathrm{L}} a_i)^3}{1+4(k_{0\mathrm{L}} a_i)^2} + \sum_i A_i \frac{4Q_{\mathrm{L}\to\mathrm{T}} \left(1+4Q_{\mathrm{L}\to\mathrm{T}} 2q \cdot p_i^{\mathrm{Im}}\right) k_{0\mathrm{L}} (k_{0\mathrm{T}} a_i)^3}{\left[1+(k_{0\mathrm{T}} a_i)^2 (\eta_{\mathrm{LT}}^2 - 1)\right]^2 + 4(k_{0\mathrm{T}} a_i)^2}, \tag{6}$$

$$\begin{aligned} \mathrm{Re} k_{\mathrm{L}} &= k_{0\mathrm{L}} + \sum_i A_i \frac{2Q_{\mathrm{L}\to\mathrm{L}} k_{0\mathrm{L}} (k_{0\mathrm{L}} a_i)^2}{1+4(k_{0\mathrm{L}} a_i)^2} + 2Q_{\mathrm{LL}}^* k_{0\mathrm{L}} \\ &+ \sum_i A_i \frac{2Q_{\mathrm{L}\to\mathrm{T}} \left(1+2Q_{\mathrm{L}\to\mathrm{T}} q \cdot p_i^{\mathrm{Re}}\right) k_{0\mathrm{L}} (k_{0\mathrm{T}} a_i)^2 \left[1+(k_{0\mathrm{T}} a_i)^2 (\eta_{\mathrm{LT}}^2 - 1)\right]}{\left[1+(k_{0\mathrm{T}} a_i)^2 (\eta_{\mathrm{LT}}^2 - 1)\right]^2 + 4(k_{0\mathrm{T}} a_i)^2}, \\ &+ \sum_i 2A_i Q_{\mathrm{L}\to\mathrm{T}} \left(1+2Q_{\mathrm{L}\to\mathrm{T}} q \cdot p_i^{\mathrm{Re}}\right) k_{0\mathrm{L}} \end{aligned} \tag{7}$$

where the corrective factors $p_i^{\mathrm{Im}} = 1\big/\left[1+(\eta_{\mathrm{LT}}^2+1)(k_{0\mathrm{T}} a_i)^2\right]$ and $p_i^{\mathrm{Re}} = 1\big/\left[1+(\eta_{\mathrm{LT}}^2+1)(k_{0\mathrm{T}} a_i)^2/2\right]$ are included to deliver the quadratic behaviour of the model in the Rayleigh regime while retaining the same frequency behaviour in the stochastic regime as the original, uncorrected model [36]. $\eta_{\mathrm{LT}} = V_{0\mathrm{T}}/V_{0\mathrm{L}}$ is the velocity ratio. $A_i$ and $a_i$ are the coefficients of the generalised spatial TPC function [36]. In [36], the coefficient $q$ was obtained by fitting the model to the FEM results at low frequencies. Accordingly, the Rayleigh attenuation and phase velocity asymptotes at the quasi-static limit are

$$\alpha_{\mathrm{L}}^{\mathrm{R}} = \frac{1}{2\pi} k_{0\mathrm{L}}^4 V_{\mathrm{eff}}^{\mathrm{g}} \left[ Q_{\mathrm{L}\to\mathrm{L}} + \frac{V_{0\mathrm{L}}^3}{V_{0\mathrm{T}}^3} \left(1+4Q_{\mathrm{L}\to\mathrm{T}} 2q\right) Q_{\mathrm{L}\to\mathrm{T}} \right], \; V_{\mathrm{L}}^{\mathrm{R}} = \frac{V_{0\mathrm{L}}}{1+2Q_{\mathrm{LL}}^* + 2Q_{\mathrm{L}\to\mathrm{T}} \left(1+2Q_{\mathrm{L}\to\mathrm{T}} q\right)}. \tag{8}$$

In the prior work [36], we made a notable observation from the FEM results that the quadratic coefficient for attenuation is twice that for velocity variation, as has already been accounted for in Equations (6) and (7). Furthermore, the quadratic coefficient is independent of the spatial TPC and therefore is independent of a specific material microstructure; the resulting FAM has demonstrated excellent applicability to different microstructures with contrasting grain size distributions and grain shapes [36,37]. Most importantly, the applicability extends exceptionally well to the quasi-static velocity limit. This is further supported by the fact that the quadratic coefficient $q$ of $\pi^3/2$ obtained in our prior work [36] is essentially the same as that from the quasi-static SC fit in Table 3 for cubic materials with $A^{\mathrm{eq}} > 1$.

In this work, we determine the quadratic coefficient of the FAM differently by obtaining it from the SC model fit of the quasi-static velocity results. We have extended the FAM to other symmetries, for which the quadratic coefficient $q$ in the Equations (6) and (7) is taken from the quasi-static SC fits in Table 3. In this extension, we also assume that the quadratic coefficient for attenuation is twice that for phase velocity variation for all symmetries (which exhibits some inaccuracy, as we will show later). The resulting FAM predictions are provided in Figure 3 for materials with different symmetries. The figure shows that:

(1) The FAM exhibits substantially better agreement with the FEM points than the classical SOA theory for all cubic and trigonal materials with $A^{\mathrm{eq}} > 1$. Such improved agreement is related to the tight alignment of the quasi-static SC results with their fits in Figure 2.

(2) The FAM performs less satisfactorily for other symmetries and branches. This is because the quadratic coefficient of the FAM is taken from the SC fits that have a significant scatter (as shown in Figure 2) for individual quasi-static SC results in these cases.

Evidently, the FAM would be unsuitable for materials whose SC results spread out over a considerable distance from the SC fits. This leads us to propose the direct approximate model (DAM) below that is valid for individual materials.

## (b) Direct approximate model (DAM)

The DAM still uses Equations (6) and (7) for attenuation and phase velocity calculations, but the quadratic coefficient $q$ is obtained separately for each individual material. To obtain the quadratic coefficient, we start from the quasi-static Equations (5) and (8) by defining $V_{\mathrm{L}}^{\mathrm{DAM}} = V_{\mathrm{L}}^{\mathrm{SOA}} - 4Q_{\mathrm{L}\to\mathrm{T}}^{2} q V_{0\mathrm{L}}$, where $V_{\mathrm{L}}^{\mathrm{DAM}}$ and $V_{\mathrm{L}}^{\mathrm{SOA}}$ are the quasi-static velocities of the DAM and SOA models. Next, we equate $V_{\mathrm{L}}^{\mathrm{DAM}} = V_{\mathrm{L}}^{\mathrm{SC}}$; $V_{\mathrm{L}}^{\mathrm{SC}}$ is obtainable for a polycrystal of any crystal symmetry and elastic properties [60,61]. The coefficient $q$ for a particular material is thus obtained theoretically from the quasi-static velocities of the SOA and the SC models by

$$q = (V_{\mathrm{L}}^{\mathrm{SC}} - V_{\mathrm{L}}^{\mathrm{SOA}}) / (-4Q_{\mathrm{L}\to\mathrm{T}}^{2} V_{0\mathrm{L}}) . \tag{9}$$

Figure 4(a) exemplifies the determination of the quadratic coefficient $q$ for $Na_2U_1O_4$ and $Sr_1Mg_6Ga_1$. The two example materials are chosen from the $A^{\mathrm{eq}} < 1$ and $A^{\mathrm{eq}} > 1$ branches of the orthorhombic symmetry, with parameters in Table 1 and quasi-static velocity variations marked as star points in Figure 2. As shown in Figure 4(a), the two materials represent two extreme cases: their quasi-static velocity variations are overestimated and underestimated by the SOA theory relative to the SC theory, resulting in negative and positive $q$ coefficients, respectively.

Figure 4(b) compares the DAM with the FEM, SOA and FAM for the two example materials. For $Na_2U_1O_4$, the DAM delivers an attenuation curve that agrees with the FEM results even worse than the SOA and FAM in the considered low-frequency range. By contrast, the DAM shows dramatically improved accuracy in calculating phase velocity for this material. We shall see that such contrasting agreement is somewhat universal for the materials whose quasi-static velocities are overestimated by the classical SOA theory (rare cases with negative values in Figure 2(b)). For $Sr_1Mg_6Ga_1$, the DAM exhibits much better agreement with the FEM than the other two models in both attenuation and phase velocity. However, the DAM still has an apparent deviation from the FEM because, for this material, the SC result (based on which the coefficient $q$ of the DAM is calculated) has a significant difference from the accurate FEM result, as can be seen in Figure 2(b).

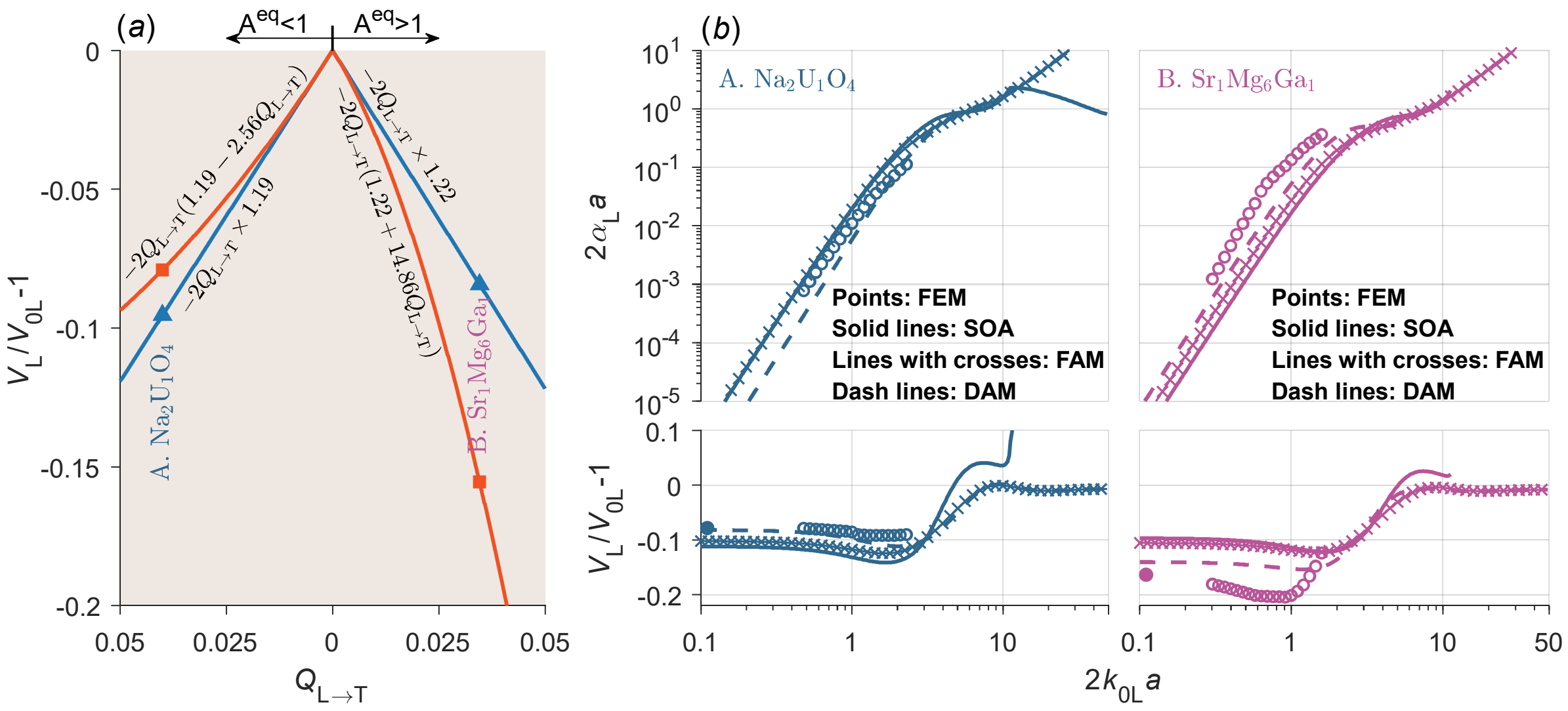

*Figure 4. (a) Example determination of the quadratic coefficient $q$ for the DAM model from the difference between the quasi-static SC (squares) and SOA (triangles) velocity variations for orthorhombic $Na_2U_1O_4$ and $Sr_1Mg_6Ga_1$. (b) shows the normalised attenuation (top) and phase velocity variation (bottom) for the two materials, comparing the FEM, SOA, FAM and DAM results.*

The above extreme examples show that the DAM generally performs better than the FAM and the classical SOA theory. This improvement is further supported by the results shown in Figure 5, a replot of Figure 3 to compare the DAM with the FEM and FAM. It can be seen that both the FAM and DAM are practically indistinguishable from the FEM results for all shown cases, justifying their suitability for predicting attenuation and phase velocity in practice. Yet, the DAM exhibits an even better (near perfect for phase velocity) agreement with the FEM than the FAM.

## (c) Quantitative evaluation of the approximate models

Now we continue to evaluate the two approximate models quantitatively. In addition to the materials used in Figures 3-5, the unhighlighted materials in Table 1 are also used for the evaluation. The attenuation and phase velocity results for these additional materials are presented in Figure 6, comparing the FEM, SOA and DAM. Note that the FEM results for these cases are given in a narrower frequency range around $2k_{0\mathrm{L}}a=1$. For the figure clarity, the FAM results are not shown in the figure, but they are quantitatively evaluated in Tables 4-5. Altogether, we have provided for the first time a rather complete attenuation and phase velocity database for 90 materials belonging to the seven crystal symmetries. For each of these materials, we use the normalised root-mean-square deviation (NRMSD) to quantify the overall difference of its SOA, FAM and DAM curves to the FEM results in the frequency range of $2k_{0\mathrm{L}}a=1$. For instance, the attenuation NRMSD of the SOA is determined by $\mathrm{rms}(\alpha_{\mathrm{L}}^{\mathrm{SOA}}/\alpha_{\mathrm{L}}^{\mathrm{FEM}}-1)_{2k_{0\mathrm{L}}a<1}$. The resulting NRMSD values for all 90 materials are listed in Table 4 and Table 5 for attenuation and phase velocity.

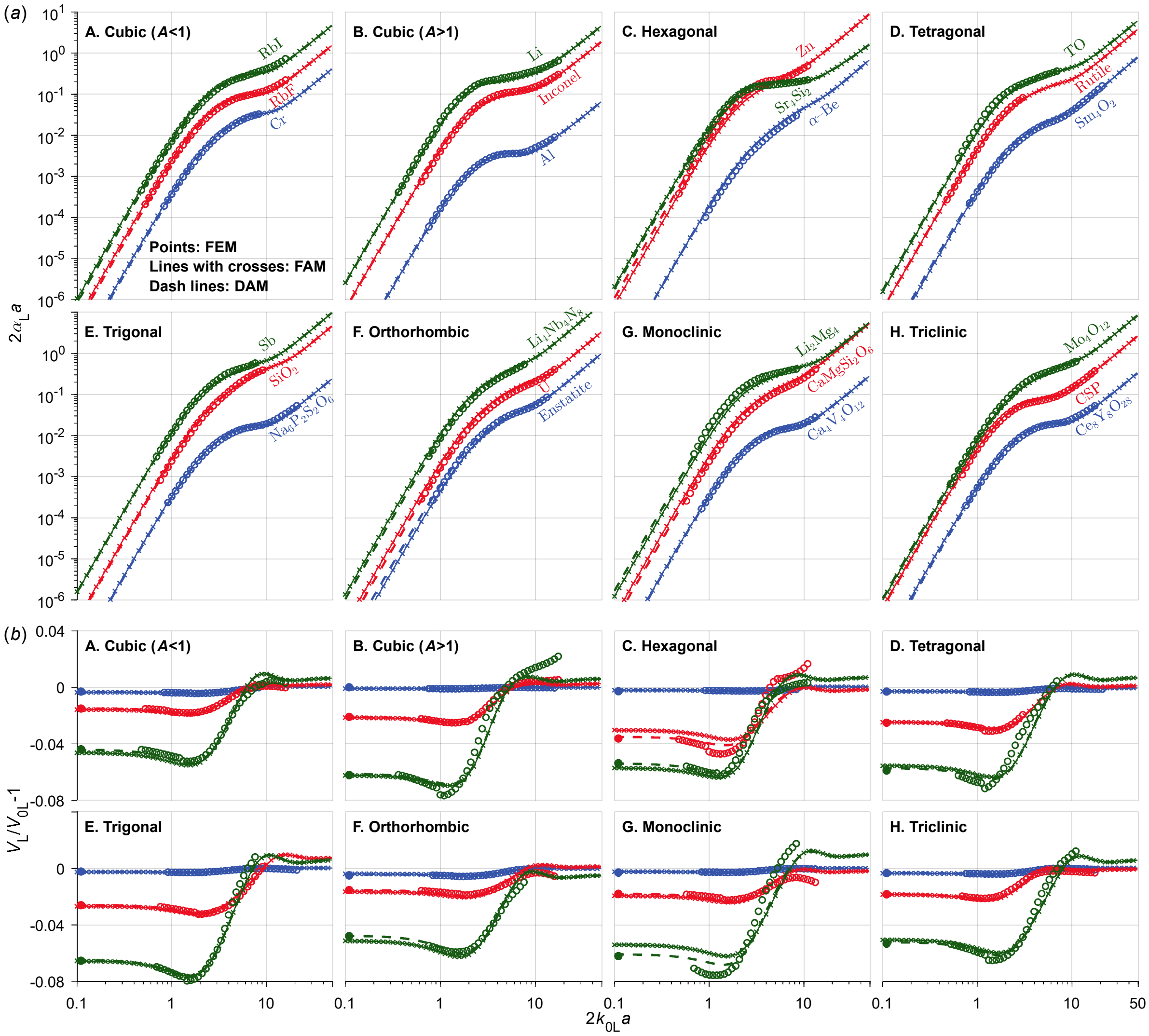

*Figure 5. Attenuation and phase velocity of plane longitudinal waves in materials of different crystal symmetries, comparing the FEM, FAM and DAM results. (a) shows the normalised attenuation and (b) the respective phase velocity variation against the normalised frequency. The FEM points and FAM curves are taken from Figure 3. The DAM curves are calculated using Equations (6) and (7), with the quadratic coefficient determined from Equation (9).*

Combining the results in Figures 3-6 and the NRMSD values in Tables 4-5, we can reach the findings below:

(1) For phase velocity, the FAM generally has a better agreement with the FEM than the classical SOA theory, and the DAM achieves a further accuracy improvement in most cases. These are especially true for both branches of cubic and hexagonal materials, the $A^{\mathrm{eq}} > 1$ branch of trigonal materials, and the $A^{\mathrm{eq}} < 1$ branch of orthorhombic materials. For these cases, the DAM has about ten times better accuracy than the SOA.

(2) For attenuation, the FAM mostly delivers an order of magnitude improvement in accuracy than the SOA, with the DAM performing even better, especially for materials of large $Q_{\mathrm{L}\to\mathrm{T}}$. However, as aforementioned, the approximate models (notably the DAM) predict even less accurate attenuation results than the SOA for the materials with a negative quadratic coefficient $q$. Such materials mainly come from the $A^{\mathrm{eq}} < 1$ branch of the cubic (of small $Q_{\mathrm{L}\to\mathrm{T}}$), hexagonal and orthorhombic symmetries, in which cases, a substantial number of materials have an SOA-SC difference below the 0% line in Figure 2(b). Since the approximate models perform exceptionally well for phase velocity, we postulate that for these cases, attenuation calculation can no longer be based on Equation (6), where the quadratic coefficient is simply the double of that determined by the quasi-static velocities. Instead, the quadratic coefficient for attenuation may have a different relation to that for phase velocity, but this is not yet understood. For this reason, the FAM and DAM should not be used for attenuation in polycrystals with negative values of $q$.

(3) For both phase velocity and attenuation, the SOA, FAM and DAM perform just as well for weakly scattering materials with $Q_{\mathrm{L}\to\mathrm{T}} \leq 0.005$. It is difficult to tell which model performs better, so the classical SOA theory would be the best choice for these materials. Such materials include widely used structural materials, such as Al, $\alpha$-Ti and $\alpha/\beta$ Ti.

We note that the FAM is empirical due to its root in the quadratic fit to the quasi-static SC velocities. By contrast, the DAM is a general model with theoretically-determined coefficients and is valid for any material. It provides a simple means for the accurate calculation of attenuation and phase velocity, especially for materials of large $Q_{\mathrm{L}\to\mathrm{T}}$. One can easily determine its needed parameter from the classical SOA and SC theories.

We emphasise that the quadratic coefficient $q$ obtained from quasi-static velocities for the FAM and DAM models is independent of the spatial TPC function. We have previously demonstrated this independence for polycrystals with equiaxed grains (scalar TPC) of greatly contrasting size distributions [36] and for polycrystals with elongated grains (direction-dependent TPC) [37]. As a result, both models presented here are applicable to any grain geometry with different grain shapes and grain size distributions.

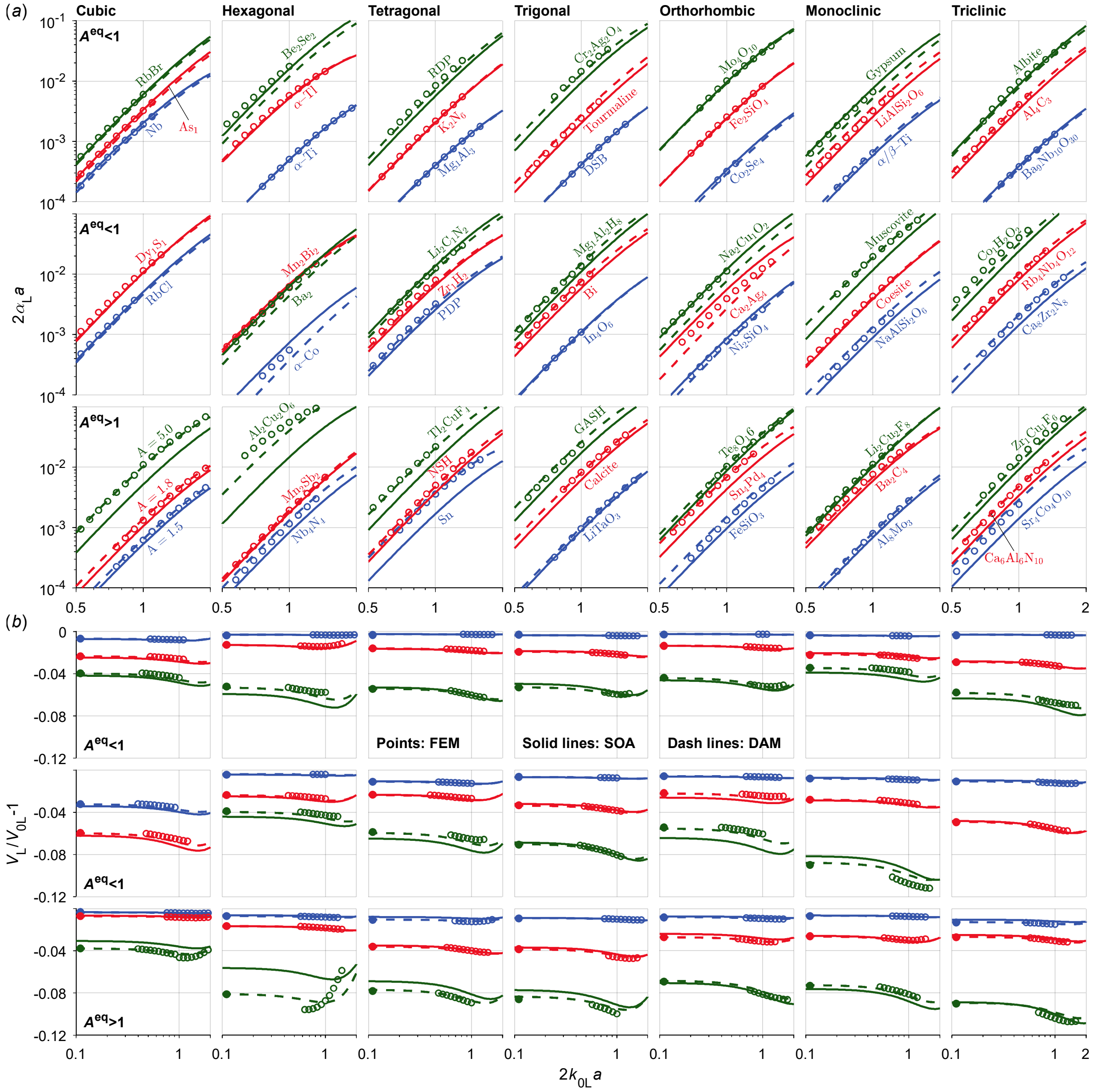

*Figure 6. Attenuation and phase velocity of plane longitudinal waves in materials of different crystal symmetries, comparing the FEM (points), SOA (solid lines) and DAM (dash lines) results. (a) shows the normalised attenuation and (b) the respective phase velocity variation against the normalised frequency. Each column corresponds to a crystal symmetry as labelled on the top of the figure. In (a) and (b), the first two rows are for materials with $A^{\mathrm{eq}} < 1$ and the third row for those with $A^{\mathrm{eq}} > 1$. Notice the different x-axis ranges used in (a) and (b).*

*Table 4. Normalised root-mean-square deviation (NRMSD) in attenuation. Each result quantifies the NRMSD of the SOA, FAM, or DAM attenuation to the FEM attenuation in the low-frequency range of $2k_{0L}a<1$. The O* row shows the two extreme orthorhombic materials. The materials with a dagger (†) superscript have negative values of $q$.*

| | $A^{eq}<1$ materials | $Q_{L\to T}$ | NRMSD to FEM (%) | | | $A^{eq}>1$ materials | $Q_{L\to T}$ | NRMSD to FEM (%) | | |
|---|---|---|---|---|---|---|---|---|---|---|
| | | | SOA | FAM | DAM | | | SOA | FAM | DAM |
| Cubic | **Cr†** | 1.40E-03 | 2.30 | 5.85 | 6.45 | **Al** | 3.34E-04 | 8.02 | 4.28 | 2.05 |
| | Nb† | 3.13E-03 | 3.29 | 6.47 | 10.22 | A=1.5 | 1.43E-03 | 14.20 | 1.26 | 4.71 |
| | **RbF†** | 6.44E-03 | 1.58 | 7.28 | 13.05 | A=1.8 | 2.79E-03 | 19.63 | 3.21 | 7.25 |
| | $As_1$† | 9.98E-03 | 1.82 | 9.61 | 11.97 | A=2.4 | 5.48E-03 | 28.98 | 8.73 | 8.73 |
| | RbCl | 1.38E-02 | 4.62 | 8.54 | 12.43 | Cu | 7.19E-03 | 36.80 | 8.01 | 8.27 |
| | RbBr | 1.67E-02 | 5.61 | 9.00 | 11.38 | **Inconel** | 7.59E-03 | 34.15 | 14.17 | 9.87 |
| | **RbI** | 1.86E-02 | 7.86 | 6.42 | 11.84 | A=5.0 | 1.27E-02 | 50.25 | 9.95 | 5.68 |
| | $Dy_1S_1$ | 2.50E-02 | 15.80 | 2.65 | 13.15 | **Li** | 1.87E-02 | 63.50 | 4.50 | 7.34 |
| Hexagonal | α-Ti | 1.22E-03 | 1.36 | 1.55 | 1.37 | **α-Be** | 7.10E-04 | 6.31 | 30.38 | 29.96 |
| | α-Co† | 1.82E-03 | 30.98 | 29.81 | 35.63 | $Nb_4N_4$ | 2.46E-03 | 25.05 | 9.54 | 21.38 |
| | α-Tl | 5.72E-03 | 13.68 | 16.58 | 8.86 | $Mn_2Sb_2$ | 6.52E-03 | 9.27 | 38.07 | 0.75 |
| | $Mn_2Bi_2$† | 9.85E-03 | 7.87 | 2.49 | 2.56 | **Zn** | 9.97E-03 | 56.86 | 22.70 | 18.51 |
| | $Ba_2$† | 1.76E-02 | 1.98 | 12.00 | 28.50 | **$Sr_4Si_2$** | 1.83E-02 | 47.60 | 13.62 | 2.74 |
| | $Be_2Se_2$† | 2.50E-02 | 20.57 | 34.57 | 40.43 | $Al_2Cu_2O_6$ | 2.34E-02 | 78.51 | 48.53 | 38.54 |
| Tetragonal | $Mg_1Al_3$† | 1.02E-03 | 1.94 | 0.29 | 1.48 | **$Sm_4O_2$** | 9.96E-04 | 18.02 | 5.16 | 8.52 |
| | PDP | 4.59E-03 | 19.92 | 16.60 | 5.04 | Sn | 3.39E-03 | 51.97 | 41.87 | 12.26 |
| | $K_2N_6$† | 6.30E-03 | 6.74 | 5.98 | 4.76 | **Rutile** | 8.70E-03 | 32.62 | 3.70 | 4.29 |
| | $Zr_1H_2$ | 9.92E-03 | 18.89 | 12.38 | 8.51 | NSH | 1.27E-02 | 22.03 | 31.58 | 5.85 |
| | RDP | 2.02E-02 | 32.93 | 24.53 | 14.16 | **TO** | 1.83E-02 | 61.78 | 25.48 | 19.40 |
| | $Li_2C_1N_2$ | 2.47E-02 | 5.81 | 18.10 | 10.90 | $Tl_2Cu_1F_4$ | 2.67E-02 | 46.96 | 23.50 | 7.48 |
| Trigonal | DSB | 1.27E-03 | 2.20 | 4.12 | 4.19 | **$Na_6P_2S_2O_6$†** | 9.83E-04 | 2.43 | 1.44 | 4.13 |
| | $In_4O_6$† | 2.56E-03 | 1.45 | 8.71 | 4.73 | $LiTaO_3$ | 3.30E-03 | 5.46 | 9.27 | 9.31 |
| | Tourmaline | 6.69E-03 | 18.82 | 5.07 | 13.34 | **$SiO_2$** | 9.58E-03 | 10.72 | 16.70 | 4.90 |
| | Bi | 1.29E-02 | 22.93 | 5.18 | 6.44 | Calcite | 1.46E-02 | 26.34 | 3.10 | 2.23 |
| | $Cr_2Ag_2O_4$ | 1.97E-02 | 36.87 | 17.83 | 11.29 | **Sb** | 2.29E-02 | 33.59 | 2.69 | 1.02 |
| | $Mg_1Al_2H_8$ | 2.48E-02 | 22.57 | 12.69 | 10.86 | GASH | 2.84E-02 | 29.57 | 13.20 | 12.34 |
| Orthorhombic | $Co_2Se_4$† | 9.93E-04 | 3.94 | 6.57 | 10.20 | **Enstatite** | 1.40E-03 | 25.18 | 14.76 | 18.13 |
| | $Ni_2SiO_4$† | 2.31E-03 | 8.04 | 20.12 | 14.59 | $FeSiO_3$ | 2.70E-03 | 24.22 | 14.29 | 21.03 |
| | $Fe_2SiO_4$ | 5.37E-03 | 4.86 | 2.20 | 3.14 | **U†** | 6.03E-03 | 5.72 | 30.03 | 5.26 |
| | $Ca_2Ag_4$† | 1.02E-02 | 40.27 | 41.89 | 36.41 | $Sn_4Pd_4$ | 9.85E-03 | 36.15 | 17.74 | 6.57 |
| | $Mo_4O_{10}$ | 1.76E-02 | 4.75 | 4.25 | 8.24 | **$Li_4Nb_4N_8$** | 1.77E-02 | 19.85 | 21.55 | 4.47 |
| | $Na_2Cu_1O_2$† | 2.55E-02 | 4.35 | 3.20 | 38.25 | $Te_8O_{16}$ | 2.43E-02 | 13.11 | 40.23 | 6.97 |
| O* | $Na_2U_1O_4$† | 4.00E-02 | 77.75 | 72.57 | 48.47 | $Sr_1Mg_6Ga_1$ | 3.46E-02 | 89.62 | 81.70 | 64.17 |
| Monoclinic | α/β Ti | 1.52E-03 | 11.29 | 5.59 | 2.40 | **$Ca_4V_4O_{12}$†** | 9.28E-04 | 1.03 | 5.69 | 0.46 |
| | $NaAlSi_2O_6$ | 2.80E-03 | 24.02 | 12.63 | 15.79 | $Al_8Mo_3$ | 2.54E-03 | 11.54 | 4.51 | 6.95 |
| | $LiAlSi_2O_6$ | 7.40E-03 | 20.54 | 1.12 | 14.44 | **$CaMgSi_2O_6$†** | 6.63E-03 | 5.76 | 42.00 | 7.57 |
| | Coesite | 1.08E-02 | 11.93 | 18.34 | 6.54 | $Ba_2C_4$ | 1.09E-02 | 24.32 | 15.89 | 11.36 |
| | Gypsum† | 1.56E-02 | 10.21 | 57.36 | 26.16 | **$Li_2Mg_4$** | 1.82E-02 | 65.71 | 38.65 | 17.86 |
| | Muscovite | 2.90E-02 | 40.24 | 3.64 | 3.54 | $Li_2Cu_2F_8$ | 2.60E-02 | 11.21 | 81.34 | 3.83 |
| Triclinic | $Ba_9Nb_{10}O_{30}$† | 1.20E-03 | 2.07 | 6.07 | 6.99 | **$Ce_8Y_8O_{28}$†** | 1.42E-03 | 0.91 | 6.81 | 7.06 |
| | $Ca_8Zr_2N_8$ | 3.96E-03 | 26.36 | 19.63 | 10.13 | $Sr_4Co_4O_{10}$ | 3.93E-03 | 38.87 | 29.28 | 30.55 |
| | $Al_4C_3$ | 1.09E-02 | 17.37 | 7.04 | 7.37 | **CSP** | 7.19E-03 | 14.38 | 8.69 | 5.43 |
| | $Rb_4Nb_4O_{12}$ | 1.78E-02 | 18.50 | 8.64 | 5.42 | $Ca_6Al_6N_{10}$ | 9.78E-03 | 27.62 | 2.76 | 4.95 |
| | Albite | 2.36E-02 | 11.18 | 27.18 | 18.78 | **$Mo_4O_{12}$** | 1.74E-02 | 36.78 | 5.35 | 6.94 |
| | $Co_1H_2O_2$ | 4.49E-02 | 45.74 | 20.39 | 24.05 | $Zr_1Cu_1F_6$ | 3.08E-02 | 34.42 | 9.84 | 14.09 |

*Table 5. Normalised root-mean-square deviation (NRMSD) in phase velocity. Each result quantifies the NRMSD of the SOA, FAM, or DAM velocity to the FEM velocity in the low-frequency range of $2k_{0\mathrm{L}}a<1$. The O* row shows the two extreme orthorhombic materials. The materials with a dagger (†) superscript have negative values of $q$.*

| | $A^{\mathrm{eq}}<1$ materials | $Q_{\mathrm{L}\to\mathrm{T}}$ | NRMSD to FEM (%) SOA | FAM | DAM | $A^{\mathrm{eq}}>1$ materials | $Q_{\mathrm{L}\to\mathrm{T}}$ | NRMSD to FEM (%) SOA | FAM | DAM |
|---|---|---|---|---|---|---|---|---|---|---|
| Cubic | **Cr†** | 1.40E-03 | 0.025 | 0.025 | 0.015 | **Al** | 3.34E-04 | 0.0002 | 0.001 | 0.002 |
| | Nb† | 3.13E-03 | 0.073 | 0.073 | 0.044 | A=1.5 | 1.43E-03 | 0.012 | 0.003 | 0.004 |
| | **RbF†** | 6.44E-03 | 0.147 | 0.144 | 0.070 | A=1.8 | 2.79E-03 | 0.039 | 0.004 | 0.010 |
| | $As_1$† | 9.98E-03 | 0.260 | 0.247 | 0.120 | A=2.4 | 5.48E-03 | 0.144 | 0.011 | 0.011 |
| | RbCl | 1.38E-02 | 0.358 | 0.338 | 0.162 | Cu | 7.19E-03 | 0.248 | 0.029 | 0.030 |
| | RbBr | 1.67E-02 | 0.446 | 0.405 | 0.194 | **Inconel** | 7.59E-03 | 0.245 | 0.049 | 0.026 |
| | **RbI** | 1.86E-02 | 0.493 | 0.436 | 0.213 | A=5.0 | 1.27E-02 | 0.750 | 0.104 | 0.093 |
| | $Dy_1S_1$ | 2.50E-02 | 0.609 | 0.492 | 0.259 | **Li** | 1.87E-02 | 1.752 | 0.176 | 0.240 |
| Hexagonal | α-Ti | 1.22E-03 | 0.016 | 0.015 | 0.016 | **α-Be** | 7.10E-04 | 0.001 | 0.006 | 0.006 |
| | α-Co† | 1.82E-03 | 0.085 | 0.083 | 0.030 | $Nb_4N_4$ | 2.46E-03 | 0.067 | 0.043 | 0.016 |
| | α-Tl | 5.72E-03 | 0.109 | 0.082 | 0.112 | $Mn_2Sb_2$ | 6.52E-03 | 0.072 | 0.233 | 0.073 |
| | $Mn_2Bi_2$† | 9.85E-03 | 0.270 | 0.168 | 0.155 | **Zn** | 9.97E-03 | 1.070 | 0.698 | 0.220 |
| | $Ba_2$† | 1.76E-02 | 0.778 | 0.445 | 0.267 | **$Sr_4Si_2$** | 1.83E-02 | 0.793 | 0.360 | 0.050 |
| | $Be_2Se_2$† | 2.50E-02 | 1.136 | 0.498 | 0.382 | $Al_2Cu_2O_6$ | 2.34E-02 | 3.225 | 1.364 | 0.692 |
| Tetragonal | $Mg_1Al_3$† | 1.02E-03 | 0.016 | 0.016 | 0.015 | **$Sm_4O_2$** | 9.96E-04 | 0.008 | 0.001 | 0.009 |
| | PDP | 4.59E-03 | 0.005 | 0.004 | 0.043 | Sn | 3.39E-03 | 0.284 | 0.249 | 0.031 |
| | $K_2N_6$† | 6.30E-03 | 0.109 | 0.109 | 0.073 | **Rutile** | 8.70E-03 | 0.200 | 0.028 | 0.030 |
| | $Zr_1H_2$ | 9.92E-03 | 0.094 | 0.083 | 0.114 | NSH | 1.27E-02 | 0.015 | 0.392 | 0.089 |
| | RDP | 2.02E-02 | 0.016 | 0.140 | 0.077 | **TO** | 1.83E-02 | 1.351 | 0.487 | 0.302 |
| | $Li_2C_1N_2$ | 2.47E-02 | 1.079 | 0.834 | 0.380 | $Tl_2Cu_1F_4$ | 2.67E-02 | 0.874 | 0.914 | 0.112 |
| Trigonal | DSB | 1.27E-03 | 0.012 | 0.014 | 0.014 | **$Na_6P_2S_2O_6$†** | 9.83E-04 | 0.010 | 0.012 | 0.008 |
| | $In_4O_6$† | 2.56E-03 | 0.045 | 0.052 | 0.032 | $LiTaO_3$ | 3.30E-03 | 0.017 | 0.031 | 0.031 |
| | Tourmaline | 6.69E-03 | 0.021 | 0.017 | 0.064 | **$SiO_2$** | 9.58E-03 | 0.076 | 0.162 | 0.079 |
| | Bi | 1.29E-02 | 0.050 | 0.037 | 0.095 | Calcite | 1.46E-02 | 0.297 | 0.094 | 0.158 |
| | $Cr_2Ag_2O_4$ | 1.97E-02 | 0.184 | 0.065 | 0.069 | **Sb** | 2.29E-02 | 0.393 | 0.033 | 0.052 |
| | $Mg_1Al_2H_8$ | 2.48E-02 | 0.017 | 0.138 | 0.094 | GASH | 2.84E-02 | 0.860 | 0.325 | 0.348 |
| Orthorhombic | $Co_2Se_4$† | 9.93E-04 | 0.022 | 0.023 | 0.013 | **Enstatite** | 1.40E-03 | 0.047 | 0.041 | 0.003 |
| | $Ni_2SiO_4$† | 2.31E-03 | 0.070 | 0.076 | 0.031 | $FeSiO_3$ | 2.70E-03 | 0.082 | 0.069 | 0.006 |
| | $Fe_2SiO_4$ | 5.37E-03 | 0.075 | 0.068 | 0.055 | **U†** | 6.03E-03 | 0.162 | 0.205 | 0.083 |
| | $Ca_2Ag_4$† | 1.02E-02 | 0.623 | 0.559 | 0.209 | $Sn_4Pd_4$ | 9.85E-03 | 0.300 | 0.195 | 0.028 |
| | $Mo_4O_{10}$ | 1.76E-02 | 0.235 | 0.017 | 0.051 | **$Li_4Nb_4N_8$** | 1.77E-02 | 0.059 | 0.226 | 0.152 |
| | $Na_2Cu_1O_2$† | 2.55E-02 | 1.444 | 1.055 | 0.400 | $Te_8O_{16}$ | 2.43E-02 | 0.013 | 0.394 | 0.279 |
| O* | $Na_2U_1O_4$† | 4.00E-02 | 4.605 | 3.092 | 0.926 | $Sr_1Mg_6Ga_1$ | 3.46E-02 | 11.628 | 10.722 | 6.390 |
| Monoclinic | α/β Ti | 1.52E-03 | 0.009 | 0.014 | 0.022 | **$Ca_4V_4O_{12}$†** | 9.28E-04 | 0.009 | 0.011 | 0.008 |
| | $NaAlSi_2O_6$ | 2.80E-03 | 0.053 | 0.039 | 0.021 | $Al_8Mo_3$ | 2.54E-03 | 0.002 | 0.017 | 0.022 |
| | $LiAlSi_2O_6$ | 7.40E-03 | 0.085 | 0.026 | 0.052 | **$CaMgSi_2O_6$†** | 6.63E-03 | 0.131 | 0.241 | 0.094 |
| | Coesite | 1.08E-02 | 0.184 | 0.303 | 0.116 | $Ba_2C_4$ | 1.09E-02 | 0.056 | 0.326 | 0.091 |
| | Gypsum† | 1.56E-02 | 0.693 | 0.938 | 0.219 | **$Li_2Mg_4$** | 1.82E-02 | 2.286 | 1.595 | 0.862 |
| | Muscovite | 2.90E-02 | 1.414 | 0.882 | 0.855 | $Li_2Cu_2F_8$ | 2.60E-02 | 0.664 | 1.827 | 0.177 |
| Triclinic | $Ba_9Nb_{10}O_{30}$† | 1.20E-03 | 0.020 | 0.020 | 0.011 | **$Ce_8Y_8O_{28}$†** | 1.42E-03 | 0.019 | 0.022 | 0.011 |
| | $Ca_8Zr_2N_8$ | 3.96E-03 | 0.086 | 0.077 | 0.017 | $Sr_4Co_4O_{10}$ | 3.93E-03 | 0.251 | 0.229 | 0.018 |
| | $Al_4C_3$ | 1.09E-02 | 0.024 | 0.080 | 0.084 | **CSP** | 7.19E-03 | 0.019 | 0.084 | 0.060 |
| | $Rb_4Nb_4O_{12}$ | 1.78E-02 | 0.075 | 0.171 | 0.125 | $Ca_6Al_6N_{10}$ | 9.78E-03 | 0.133 | 0.017 | 0.048 |
| | Albite | 2.36E-02 | 0.541 | 0.679 | 0.095 | **$Mo_4O_{12}$** | 1.74E-02 | 0.492 | 0.185 | 0.055 |
| | $Co_1H_2O_2$ | 4.49E-02 | 1.230 | 1.335 | 1.577 | $Zr_1Cu_1F_6$ | 3.08E-02 | 0.437 | 0.263 | 0.558 |

# 6. Conclusion

In this paper, we appraised the classical SOA theory using 3D grain-scale FE simulations, based on which we proposed approximate models to deliver more accurate attenuation and phase velocity calculations. We focused on plane longitudinal waves in untextured polycrystals with statistically equiaxed grains covering all seven crystal symmetries.

We initially appraised the classical SOA theory at the quasi-static velocity limit using the 3D FE and SC results. We revealed prominent findings for the addressed longitudinal velocity limit $V_{\mathrm{L}}^{\mathrm{SOA}}$ based on 10,390 materials belonging to the seven crystal symmetries. First, the SC ( $V_{\mathrm{L}}^{\mathrm{SC}}$ ) and FEM quasistatic velocities have an excellent agreement (error below ~0.1%), so the SC theory is suited for appraising the SOA theory at the quasi-static velocity limit. Second, the SC and FEM results generally exhibit two branches with $A^{\mathrm{eq}} < 1$ and $A^{\mathrm{eq}} > 1$ ( $A^{\mathrm{eq}} = 1$ denotes isotropy) for each crystal symmetry. Third, the SC and FEM results of each branch are better sorted by the elastic scattering factor, $Q_{\mathrm{L}\to\mathrm{T}}$, than by existing anisotropy indices, manifested as a monotonic correlation of the results with $Q_{\mathrm{L}\to\mathrm{T}}$. Lastly, the $V_{\mathrm{L}}^{\mathrm{SOA}}$ shows a linear relationship with $Q_{\mathrm{L}\to\mathrm{T}}$ for all symmetries and branches (since the SOA solution is truncated to this order), while by contrast, the SC and FEM velocities generally have a quadratic relationship. This led us to postulate that the quasi-static velocity variation of any material can be described up to the quadratic order of $Q_{\mathrm{L}\to\mathrm{T}}$, with the linear part defined by the SOA theory and the quadratic part by the $V_{\mathrm{L}}^{\mathrm{SC}}$ - $V_{\mathrm{L}}^{\mathrm{SOA}}$ difference.

We subsequently appraised the classical SOA theory for its goodness in calculating attenuation and phase velocity using the FEM results of 90 materials with different crystal symmetries. We found that the SOA theory has an excellent agreement with the FEM results for materials of low $Q_{\mathrm{L}\to\mathrm{T}}$, but it shows an increasingly large deviation from the FEM as $Q_{\mathrm{L}\to\mathrm{T}}$ increases. Following the quasi-static velocity findings, we attributed this rise in deviation to the different orders of the SOA and the FEM on $Q_{\mathrm{L}\to\mathrm{T}}$, with the former being a linear order and the latter quadratic. Based on these results, we proposed the fitted (FAM) and direct (DAM) approximate models to account for the additional quadratic term in the models. The FAM takes its quadratic coefficient from the quasi-static $V_{\mathrm{L}}^{\mathrm{SC}}$ fits and is thus symmetry-specific. Its predictions are mostly indistinguishable from the FEM results, but it delivers excellent results only for cubic and trigonal materials with $A^{\mathrm{eq}} > 1$ whose quasi-static SC results follow tightly with the fits. The DAM takes a step further to theoretically determine its quadratic coefficient from the quasi-static $V_{\mathrm{L}}^{\mathrm{SC}}$ - $V_{\mathrm{L}}^{\mathrm{SOA}}$ velocity difference for each individual material. As a result, the DAM is a general model suitable for any material and has a generally better agreement with the FEM than the FAM. It exhibits a near-perfect performance for phase velocity, but for attenuation, it is only valid for materials of positive quadratic coefficients (also applies to the FAM). We do not yet understand why the models are not well suited for materials of negative quadratic coefficients. We note that the FAM and DAM models have the same level of accuracy as the classical SOA theory for weakly scattering materials with $Q_{\mathrm{L}\to\mathrm{T}} \leq 0.005$.

The proposed models (particularly the DAM) provide a simple means for accurately calculating attenuation and phase velocity. Notably, they are independent of the spatial TPC and can thus be used for various polycrystals with different grain size distributions and shapes. Therefore, they may open up exciting opportunities for characterising the microstructure of polycrystals.

# References

[1]. Aki K, Richards PG. Quantitative seismology. 2nd ed. Mill Valley, California: University Science Books; 2009.

[2]. Sato H, Fehler MC, Maeda T. Seismic Wave Propagation and Scattering in the Heterogeneous Earth. 2nd ed. Berlin: Springer; 2012.

[3]. Thompson RB. Elastic-wave propagation in random polycrystals: fundamentals and application to nondestructive evaluation. In: Fink M, Kuperman WA, Montagner J-P, Tourin A, editors. Topics in Applied Physics Imaging of complex media with acoustic and seismic waves. Berlin, Germany: Springer-Verlag; 2002. p. 233–57.

[4]. Thompson RB, Margetan FJ, Haldipur P, Yu L, Li A, Panetta PD, et al. Scattering of elastic waves in simple and complex polycrystals. Wave Motion. 2008;45(5):655–74.

[5]. Liu Y, Kalkowski MK, Huang M, Lowe MJS, Samaitis V, Cicenas V, et al. Can ultrasound attenuation measurement be used to characterise grain statistics in castings? Ultrasonics. 2021;115:106441.

[6]. Rokhlin SI, Sha G, Li J, Pilchak AL. Inversion methodology for ultrasonic characterization of polycrystals with clusters of preferentially oriented grains. Ultrasonics. 2021;115:106433.

[7]. Wydra A, Chertov AM, Maev RG, Kube CM, Du H, Turner JA. Grain Size Measurement of Copper Spot Welding Caps Via Ultrasonic Attenuation and Scattering Experiments. Res Nondestruct Eval. 2015;26(4):225–43.

[8]. Lobkis OI, Rokhlin SI. Characterization of polycrystals with elongated duplex microstructure by inversion of ultrasonic backscattering data. Appl Phys Lett. 2010 Apr 19;96(16):161905.

[9]. Shahjahan S, Aubry A, Rupin F, Chassignole B, Derode A. A random matrix approach to detect defects in a strongly scattering polycrystal: How the memory effect can help overcome multiple scattering. Appl Phys Lett. 2014;104(23).

[10]. Van Pamel A, Brett CR, Lowe MJS. A methodology for evaluating detection performance of ultrasonic array imaging algorithms for coarse-grained materials. IEEE Trans Ultrason Ferroelectr Freq Control. 2014 Dec;61(12):2042–53.

[11]. Mason WP, McSkimin HJ. Attenuation and Scattering of High Frequency Sound Waves in Metals and Glasses. J Acoust Soc Am. 1947;19(3):464–73.

[12]. Huntington HB. On Ultrasonic Scattering by Polycrystals. J Acoust Soc Am. 1950;22(3):362–4.

[13]. Mason WP, McSkimin HJ. Energy Losses of Sound Waves in Metals Due to Scattering and Diffusion. J Appl Phys. 1948 Oct;19(10):940–6.

[14]. Papadakis EP. Ultrasonic attenuation caused by scattering in polycrystalline media. In: Mason WP, editor. Physical Acoustics. New York and London: Academic Press; 1968. p. 269–328.

[15]. Stanke FE, Kino GS. A unified theory for elastic wave propagation in polycrystalline materials. J Acoust Soc Am. 1984;75(3):665–81.

[16]. Keller JB. Stochastic equations and wave propagation in random media. In: Bellman R, editor. Proceedings of Symposia in Applied Mathematics Stochastic Processes in Mathematical Physics and Engineering. Providence, RI: American Mathematical Society; 1964. p. 145–70.

[17]. Karal FC, Keller JB. Elastic, electromagnetic, and other waves in a random medium. J Math Phys. 1964;5(4):537–47.

[18]. Weaver RL. Diffusivity of ultrasound in polycrystals. J Mech Phys Solids. 1990;38(1):55–86.

[19]. Frisch U. Wave propagation in random media. In: Bharucha-Reid A, editor. Probabilistic Methods in Applied Mathematics. Academic Press, New York; 1968. p. 75–198.

[20]. Rytov SM, Kravtsov YA, Tatarskii VI. Principles of statistical radiophysics 4: wave propagation through random media. New York: Springer; 2011.

[21]. Bourret RC. Stochastically perturbed fields, with applications to wave propagation in random media. Nuovo Cim Ser 10. 1962;26(1):1–31.

[22]. Turner JA. Elastic wave propagation and scattering in heterogeneous anisotropic media: Textured polycrystalline materials. J Acoust Soc Am. 1999;106(2):541–52.

[23]. Bhattacharjee A, Pilchak AL, Lobkis OI, Foltz JW, Rokhlin SI, Williams JC. Correlating ultrasonic attenuation and microtexture in a near-alpha titanium alloy. Metall Mater Trans A. 2011;42(8):2358–72.

[24]. Calvet M, Margerin L. Velocity and attenuation of scalar and elastic waves in random media: A spectral function approach. J Acoust Soc Am. 2012;131(3):1843–62.

[25]. Kube CM. Iterative solution to bulk wave propagation in polycrystalline materials. J Acoust Soc Am. 2017;141(3):1804–11.

[26]. Shahjahan S, Rupin F, Aubry A, Chassignole B, Fouquet T, Derode A. Comparison between experimental and 2-D numerical studies of multiple scattering in Inconel600® by means of array probes. Ultrasonics. 2014;54(1):358–67.

[27]. Van Pamel A, Brett CR, Huthwaite P, Lowe MJS. Finite element modelling of elastic wave scattering within a polycrystalline material in two and three dimensions. J Acoust Soc Am. 2015;138(4):2326–36.

[28]. Van Pamel A, Sha G, Rokhlin SI, Lowe MJS. Finite-element modelling of elastic wave propagation and scattering within heterogeneous media. Proc R Soc A Math Phys Eng Sci. 2017;473(2197):20160738.

[29]. Van Pamel A, Sha G, Lowe MJS, Rokhlin SI. Numerical and analytic modelling of elastodynamic scattering within polycrystalline materials. J Acoust Soc Am. 2018;143(4):2394–408.

[30]. Ryzy M, Grabec T, Sedlák P, Veres IA. Influence of grain morphology on ultrasonic wave attenuation in polycrystalline media. J Acoust Soc Am. 2018;143(1):219–29.

[31]. Bai X, Tie B, Schmitt J-H, Aubry D. Finite element modeling of grain size effects on the ultrasonic microstructural noise backscattering in polycrystalline materials. Ultrasonics. 2018;87:182–202.

[32]. Sha G, Huang M, Lowe MJS, Rokhlin SI. Attenuation and velocity of elastic waves in polycrystals with generally anisotropic grains: analytic and numerical modeling. J Acoust Soc Am. 2020;147(4):2422–65.

[33]. Huang M, Sha G, Huthwaite P, Rokhlin SI, Lowe MJS. Elastic wave velocity dispersion in polycrystals with elongated grains: theoretical and numerical analysis. J Acoust Soc Am. 2020;148(6):3645–3662.

[34]. Huang M, Sha G, Huthwaite P, Rokhlin SI, Lowe MJS. Longitudinal wave attenuation in polycrystals with elongated grains: 3D numerical and analytical modeling. J Acoust Soc Am. 2021;149(4):2377–94.

[35]. Huang M, Sha G, Huthwaite P, Rokhlin SI, Lowe MJS. Maximizing the accuracy of finite element simulation of elastic wave propagation in polycrystals. J Acoust Soc Am. 2020;148(4):1890–910.

[36]. Huang M, Huthwaite P, Rokhlin SI, Lowe MJS. Finite element and semi-analytical study of elastic wave propagation in strongly scattering polycrystals. Proc R Soc A Math Phys Eng Sci. 2022;in print:(preprint: https://arxiv.org/abs/2111.14913).

[37]. Huang M, Rokhlin SI, Lowe MJS. Finite element evaluation of a simple model for elastic waves in strongly scattering elongated polycrystals. JASA Express Lett. 2021;1(6):064002.

[38]. Cowan ML, Beaty K, Page JH, Liu Z, Sheng P. Group velocity of acoustic waves in strongly scattering media: Dependence on the volume fraction of scatterers. Phys Rev E. 1998;58(5):6626–36.

[39]. Gower AL, Parnell WJ, Abrahams ID. Multiple Waves Propagate in Random Particulate Materials. SIAM J Appl Math. 2019;79(6):2569–92.

[40]. Hirsekorn S. Directional dependence of ultrasonic propagation in textured polycrystals. J Acoust Soc Am. 1986;79(5):1269–79.

[41]. Rokhlin SI, Li J, Sha G. Far-field scattering model for wave propagation in random media. J Acoust Soc Am. 2015;137(5):2655–69.

[42]. Norouzian M, Turner JA. Ultrasonic wave propagation predictions for polycrystalline materials using three-dimensional synthetic microstructures: Attenuation. J Acoust Soc Am. 2019;145(4):2181–91.

[43]. Quey R, Dawson PR, Barbe F. Large-scale 3D random polycrystals for the finite element method: Generation, meshing and remeshing. Comput Methods Appl Mech Eng. 2011;200(17–20):1729–45.

[44]. Man C-S, Paroni R, Xiang Y, Kenik EA. On the geometric autocorrelation function of polycrystalline materials. J Comput Appl Math. 2006;190(1):200–10.

[45]. Ranganathan SI, Ostoja-Starzewski M. Scaling function, anisotropy and the size of RVE in elastic random polycrystals. J Mech Phys Solids. 2008;56(9):2773–91.

[46]. Nygårds M. Number of grains necessary to homogenize elastic materials with cubic symmetry. Mech Mater. 2003;35(11):1049–57.

[47]. Huthwaite P. Accelerated finite element elastodynamic simulations using the GPU. J Comput Phys. 2014;257:687–707.

[48]. Sha G, Rokhlin SI. Universal Scaling of Transverse Wave Attenuation in Polycrystals. Ultrasonics. 2018;88:84–96.

[49]. Ranganathan SI, Ostoja-Starzewski M. Universal elastic anisotropy index. Phys Rev Lett. 2008;101(5):3–6.

[50]. Kube CM. Elastic anisotropy of crystals. AIP Adv. 2016 Sep;6(9):095209.

[51]. Ledbetter H, Kim S. Monocrystal elastic constants and derived properties of the cubic and the hexagonal elements. In: Levy M, Bass HE, Stern RR, editors. Handbook of Elastic Properties of Solids, Liquids and Gases Volume II. New York: Academic Press; 2001. p. 97–106.

[52]. Chung DH, Buessem WR. The Elastic Anisotropy of Crystals. J Appl Phys. 1967;38(5):2010–2.

[53]. De Jong M, Chen W, Angsten T, Jain A, Notestine R, Gamst A, et al. Charting the complete elastic properties of inorganic crystalline compounds. Sci Data. 2015;2:1–13.

[54]. Yang L, Lobkis OI, Rokhlin SI. An integrated model for ultrasonic wave propagation and scattering in a polycrystalline medium with elongated hexagonal grains. Wave Motion. 2012;49(5):544–60.

[55]. Simmons G, Wang H. Single crystal elastic constants and calculated aggregate properties: a handbook. Second. Cambridge, Massachusetts: The MIT Press; 1971.

[56]. Isaak DG. Elastic Properties of Minerals and Planetary Objects. In: Levy M, Bass HE, Stern RR, editors. Handbook of Elastic Properties of Solids, Liquids, and Gases Volume III. New York: Academic Press; 2001. p. 325–76.

[57]. Fisher ES. Temperature dependence of the elastic moduli in alpha uranium single crystals, part iv (298° to 923° K). J Nucl Mater. 1966;18(1):39–54.

[58]. Kim J-Y, Rokhlin SI. Determination of elastic constants of generally anisotropic inclined lamellar structure using line-focus acoustic microscopy. J Acoust Soc Am. 2009;126(6):2998–3007.

[59]. Brown JM, Abramson EH, Angel RJ. Triclinic elastic constants for low albite. Phys Chem Miner. 2006;33(4):256–65.

[60]. Kube CM, de Jong M. Elastic constants of polycrystals with generally anisotropic crystals. J Appl Phys. 2016;120(16):165105.

[61]. Kroner E. Berechnung der elastischen Konstanten des Vielkristalls aus den Konstanten des Einkristalls. Zeitschrift fur Phys. 1958;151(4):504–18.

[62]. Hashin Z, Shtrikman S. On some variational principles in anisotropic and nonhomogeneous elasticity. J Mech Phys Solids. 1962;10(4):335–42.

[63]. Hashin Z, Shtrikman S. A variational approach to the theory of the elastic behaviour of polycrystals. J Mech Phys Solids. 1962;10(4):343–52.

[64]. Murshed MR, Ranganathan SI. Hill–Mandel condition and bounds on lower symmetry elastic crystals. Mech Res Commun. 2017;81:7–10.

[65]. Jain A, Ong SP, Hautier G, Chen W, Richards WD, Dacek S, et al. Commentary: The materials project: A materials genome approach to accelerating materials innovation. APL Mater. 2013;1(011002).

[66]. Mouhat F, Coudert FX. Necessary and sufficient elastic stability conditions in various crystal systems. Phys Rev B. 2014;90(22):4–7.

[67]. Gibbons JD, Chakraborti S. Nonparametric Statistical Inference. 5th ed. Boca Raton, FL: CRC Press; 2011.

[68]. Kleinbaum DG, Kupper LL, Nizam A, Rosenberg ES. Applied Regression Analysis and Other Multivariable Methods. 5th ed. Boston, MA: Cengage Learning; 2014. 120–121 p.

[69]. Paszkiewicz T, Wolski S. Elastic properties of cubic crystals: Every's versus Blackman's diagram. J Phys Conf Ser. 2008;104(1):0–10.

[70]. Duffy TS. Single-crystal elastic properties of minerals and related materials with cubic symmetry. Am Mineral. 2018;103(6):977–88.

[71]. Lethbridge ZAD, Walton RI, Marmier ASH, Smith CW, Evans KE. Elastic anisotropy and extreme Poisson's ratios in single crystals. Acta Mater. 2010;58(19):6444–51.

[72]. Lebensohn RA, Liu Y, Castañeda PP. On the accuracy of the self-consistent approximation for polycrystals: Comparison with full-field numerical simulations. Acta Mater. 2004;52(18):5347–61.